%% file: activeERMain.tex
\documentclass[sigconf]{acmart}

\AtBeginDocument{%
	\providecommand\BibTeX{{%
			\normalfont B\kern-0.5em{\scshape i\kern-0.25em b}\kern-0.8em\TeX}}}

\setcopyright{acmcopyright}
\copyrightyear{2024}
\acmYear{2024}
\acmDOI{XXXXXXX.XXXXXXX}

\acmConference[]{}{}{}
\acmPrice{}
\acmISBN{}

\usepackage{amsmath}

\usepackage{amssymb}
\usepackage{amsfonts}
\usepackage{algorithmic}
\usepackage{graphicx}
\usepackage{textcomp}
\usepackage{xcolor}
\usepackage{colortbl}
\usepackage{booktabs}
\usepackage{paralist}
\usepackage{multirow}
\usepackage{subcaption}
\usepackage{tikz}
\usepackage{enumitem}
\usepackage{url}
\usepackage{balance}
\usepackage{multicol}
\usepackage{makecell}
\usepackage{arydshln}
\usepackage{graphicx}
\usepackage{comment}
\usepackage{mathtools}
\usepackage{tabularx}
\usepackage{booktabs}
\usepackage[T1]{fontenc}

\DeclarePairedDelimiter\floor{\lfloor}{\rfloor}

\def\BibTeX{{\rm B\kern-.05em{\sc i\kern-.025em b}\kern-.08em
		T\kern-.1667em\lower.7ex\hbox{E}\kern-.125emX}}

\newtheorem{example}{Example}

\usepackage[most]{tcolorbox}
\usetikzlibrary{shadows}

\newcommand{\bg}[1]{\textcolor{blue}{BG: {#1}}} 
\newcommand{\rs}[1]{\textcolor{red}{RS: {#1}}} 

\newcommand{\tup}{tuple }
\newcommand{\tups}{tuples }

\newcommand{\Kasai}{Kasai}
\newcommand{\KasaiDA}{Kasai DA}
\newcommand{\nodeweight}{\phi}
\newcommand{\edgeweight}{\pi}

\newcommand{\graphs}{pair representation graphs }

\newcommand{\graphnospace}{pair graph}

\newcommand{\Graphsnospace}{Pair graphs}
\newcommand{\update}[1]{\textcolor{black}{{#1}}} 

\begin{document}

\balance
\title{The Battleship Approach to the Low Resource Entity Matching Problem}
\author{Bar Genossar}
\affiliation{%
	\institution{Technion -- Israel Institute of Technology}
	\city{Haifa}
	\state{Israel}
}
\email{sbargen@campus.technion.ac.il}

\author{Avigdor Gal}
\affiliation{%
	\institution{Technion -- Israel Institute of Technology}
	\city{Haifa}
	\state{Israel}
}
\email{avigal@technion.ac.il}

\author{Roee Shraga}
\affiliation{%
	\institution{Worcester Polytechnic Institute}
	\city{Worcester}
	\country{Massachusetts, USA}
}
\email{rshraga@wpi.edu}

\begin{abstract}
Entity matching, a core data integration problem, is the task of deciding whether two data \tups refer to the same real-world entity. Recent advances in deep learning methods, using pre-trained language models, were proposed for resolving entity matching. Although demonstrating unprecedented results, these solutions suffer from a major drawback as they require large amounts of labeled data for training, and, as such, are inadequate to be applied to \emph{low resource entity matching} problems.
To overcome the challenge of obtaining sufficient labeled data we offer a new \emph{active learning} approach, focusing on a selection mechanism that exploits unique properties of entity matching. 
We argue that a distributed representation of a tuple pair indicates its informativeness when considered among other pairs. This is used consequently in our approach that iteratively utilizes space-aware considerations.
Bringing it all together, we treat the low resource entity matching problem as a Battleship game, hunting indicative samples, focusing on positive ones, through awareness of the latent space along with careful planning of next sampling iterations.
An extensive experimental analysis shows that the proposed algorithm outperforms state-of-the-art active learning solutions to low resource entity matching, and although using less samples, can be as successful as state-of-the-art fully trained known algorithms.
\end{abstract} 

\maketitle

\input{introduction}
\input{preliminaries}

\input{model}

\input{evaluation}

\input{experiments_results}

\input{components_analysis}
\input{related}

\input{conclusions}
\input{acknowledgments}


\newpage
\bibliographystyle{ACM-Reference-Format}
\bibliography{bibfile.bib}

\balance
\end{document}
\endinput

%% file: introduction.tex
\section{Introduction}
\label{sec:intro}
Entity matching, with variations as entity resolution, record linkage, record deduplication and more, is a core data integration task that plays a crucial role in any data project life cycle~\cite{getoor2005link, christen2012data}. Essentially, entity matching aims to identify dataset \tups that refer to the same real-world entities.


Traditional methods for entity matching and its variations focused on string similarity~\cite{levenshtein1966binary,jaro1989advances,lin1998information,jaro1995probabilistic} \update{and probabilistic approaches~\cite{fellegi1969theory}}, followed by rule-based methods~\cite{singla2006entity,singh2017synthesizing}. Learning-based approaches were also suggested~\cite{bilenko2003adaptive,konda2016magellan} with an increased focus on deep learning in recent years~\cite{joty2018distributed,mudgal2018deep,fu2019end,kasai2019low,zhao2019auto,fu2020hierarchical,li2020grapher}, culminating in the use of pre-trained language models, and specifically BERT~\cite{devlin:2018bert}, fine-tuned for entity matching~\cite{li2020deep,brunner2020entity, peeters2021dual, li2021deep}.

Machine learning approaches to entity matching are by-and-large supervised, requiring labeled data for training, and in the case of deep learning, massive amounts of them~\cite{tan2018survey, long2016unsupervised,primpeli2019wdc}. Obtaining labeled data is challenging, mostly due to the need for intensive human expert labor that accompanies the labeling process~\cite{long2015learning, long2016unsupervised, 10.1145/3546930.3547496}. This, in turn, calls for methods that reduce the labeling process load while maintaining an accurate trained matcher.

\begin{figure}[b]
	\vskip-.14in
	\centering
	\begin{subfigure}[b]{0.23\textwidth}
		\centering
		\includegraphics[width=\textwidth]{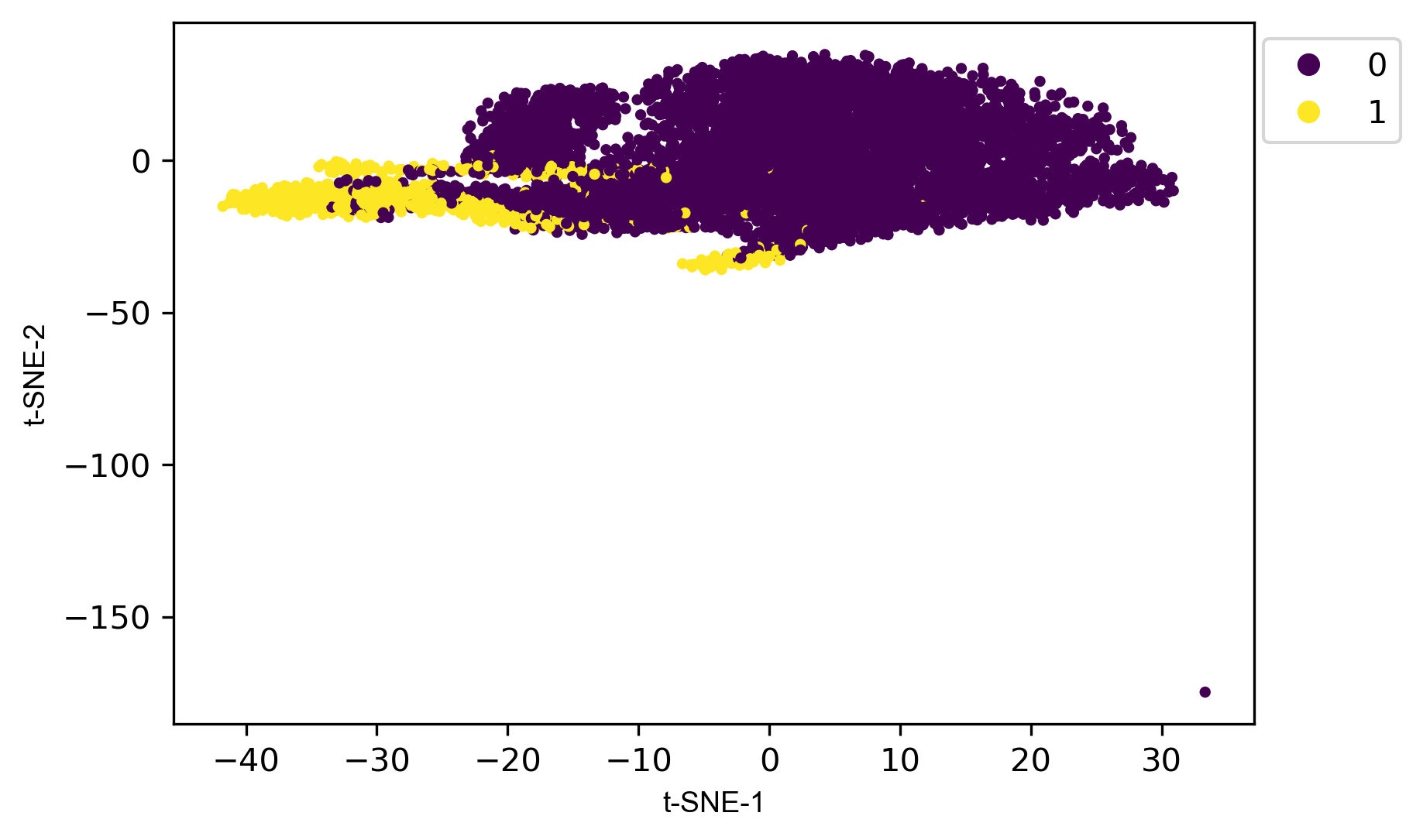}
		\caption{Amazon-Google}
		\label{fig:t-SNE_Amazon_Google}
	\end{subfigure}
	\hfill
	\begin{subfigure}[b]{0.23\textwidth}
		\centering
		\includegraphics[width=\textwidth]{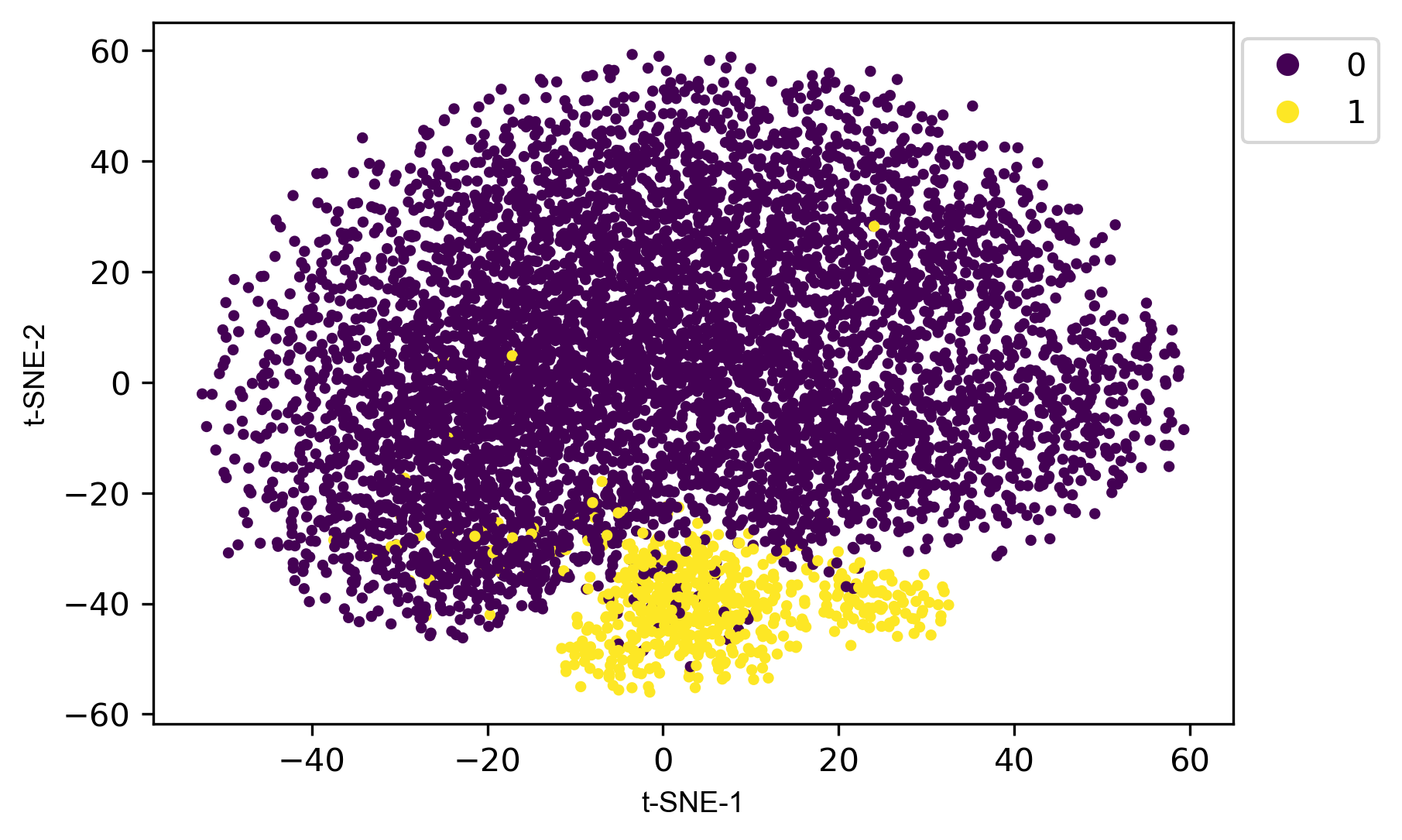}
		\caption{Walmart-Amazon}
		\label{fig:t-SNE_Walmart_Amazon}
	\end{subfigure}
	\vskip-.08in
	\caption{Visualization of pairs distribution by t-SNE, partitioned into match and non-match pairs. Using representative vectors (with dimension of 768) of a fully trained models we demonstrate that positive pairs tend to gather together.}
	\label{fig:tSNE_Plots}
\end{figure}

Several approaches were suggested over the years to overcome the dependence on large amounts of data, among which, unsupervised~\cite{wu2020zeroer, bhattacharya2006latent, zhang2020unsupervised} and active learning methods~\cite{settles2009active, prince2004does}.
Active learning, the most prevalent approach to the insufficient labeled data problem, offers methods for training using a limited amount of labeled data. This is done by zeroing in on informative samples to be labeled and used for training. An active learning framework uses selection criteria ({\em e.g.}, variations of uncertainty~\cite{meduri2020comprehensive, kasai2019low, qian2017active} and centrality~\cite{zhang2021alg, zhu2008active}) to pinpoint the samples that are expected to improve the model performance the most. These selected samples are then sent to a labeling oracle and added to the available training set in following iterations. 

A key tool which we use to quantify both uncertainty and centrality of a sample (a record pair in our case) is its vector-space latent representation. Traditional entity matching approaches constructed pair feature vector using widely used similarity measures~\cite{konda2016magellan, bilenko2003adaptive, wu2020zeroer} or learnable attribute-wise similarity vector using neural networks~\cite{mudgal2018deep, joty2018distributed}. Contemporary solutions~\cite{li2020deep,brunner2020entity, peeters2021dual}, which demonstrate superior performance and do not require feature engineering, yield an embedding vector as alternative to feature vectors when representing the input (pair) with respect to the task. 

In this work we offer a solution to low resource entity matching tasks, where there is a limited budget for labeling. In particular, we attend to the labeled data scarcity problem using active learning, proposing a novel strategy tailored to the specific characteristics of entity matching. Entity matching usually suffers from inherent label imbalance, where the number of negative (non-match) pairs overshadows the number of positive (match) ones~\cite{bellare2012active}. When training with insufficient amount of data, this property is often problematic since the model may struggle to generalize to unseen minority class samples (in this case, matching pairs). To tackle this challenge, we ensure a balanced sampling of positive and negative samples using high-dimensional latent space encoding for \tup pairs. Specifically, we argue that vector-space similarity between pair representations reflects, with high probability, agreement of their labels. We next use this notion to carefully guarantee a balanced sampling, although positive pairs are harder to be found.

\begin{example}
Figure~\ref{fig:tSNE_Plots} provides an illustration of the latent space of \tup pairs over well known benchmarks using t-SNE~\cite{van2008visualizing}. Figure~\ref{fig:t-SNE_Amazon_Google} presents the widely used Amazon-Google dataset,\footnote{\url{https://pages.cs.wisc.edu/~anhai/data1/deepmatcher_data/Structured/Amazon-Google/exp_data/}} which consists of 11,460 pairs, with only 1,167 ($\sim$$10 \%$) labeled as match samples. To demonstrate the pairs behavior, we trained a DITTO~\cite{li2020deep} model with the fully available train set ($60\%$ of the entire dataset). After training, we extracted the latent representations (dimension of 768) generated for each \tup pair 
and applied t-SNE to create a human-readable 2-dimensional reduced space, where the yellow dots represent match pairs and the purple dots represent non-match pairs.  As illustrated, there is a concentration of match pairs in a few main areas of the latent space, forming a cohesive and clear partition between match and non-match samples. As another example, Figure ~\ref{fig:t-SNE_Walmart_Amazon} presents the pairs distribution over the Walmart-Amazon dataset\footnote{\url{https://pages.cs.wisc.edu/~anhai/data1/deepmatcher_data/Structured/Walmart-Amazon/exp_data/}} (10,242 pairs, 962 matches, with $60\%$ of the entire dataset used as a train set). Again, we see a concentration of match pairs, this time mainly in a single area of the reduced latent space. 
\end{example}

This example suggests that a latent space can identify regions where mostly match or non-match samples reside. We \update{use} this observation to tackle a major problem in active learning, namely modeling \update{prediction uncertainty}. A standard approach is to use the model prediction confidence score as a proxy to its underlying uncertainty. However, when using transformer-based models this method is likely to fail, as they tend to produce extreme confidence values (close to 0 or 1) which barely reflect the real confidence~\cite{guo2017calibration, jiang2021can}. We propose a new approach for modeling the uncertainty of a pair, based on its agreement with other pairs in its vicinity, such that \update{when the correspondence between pair's prediction and its surroundings is higher, its uncertainty becomes lower.}

\subsection{Contributions}
In this work we offer a novel active learning approach to entity matching, utilizing an effective space partitioning for diversity, and locality-based measurements for targeted sample selection. Specifically, equipped with the observed phenomenon of concentration of match pairs (Figure~\ref{fig:tSNE_Plots}), we propose a sampling strategy that can be intuitively demonstrated using the popular Battleship game\footnote{\url{https://en.wikipedia.org/wiki/Battleship\_(game)}} (hence the battleship approach in the manuscript title). Our strategy uses a locality principle, searching in a vicinity of a match (or non-match) pair to create a balanced training set for human labeling. To balance locality with diversity and supporting model generalization we use a constrained version of the well-known k-means clustering algorithm~\cite{bradley2000constrained}, combined with centrality and uncertainty measures that offers an efficient selection mechanism of samples.


Our main contribution can be summarized as follows:
\begin{compactenum}
	\item A novel active learning sample selection method, based on match and non-match pair locality to ensure a balanced training set. The proposed solution uses \tup pair representation as a tool for diverse sampling.
	\item A novel uncertainty measure, overcoming the barrier of dichotomous confidence values assigned by transformer pre-trained language model.
	\item A large-scale empirical evaluation showing the effectiveness of our approach.  
	\item An open source access to our implementation.\footnote{\url{https://github.com/BarGenossar/The-Battleship-Approach-to-AL-of-EM-Problem}}
\end{compactenum}

The rest of the paper is organized as follows. We survey the necessary background for our task in Section~\ref{sec:preliminaries}, positioning entity matching in the context of active learning. In Section~\ref{sec:model} we propose a new active learning algorithm for entity matching, inspired by the battleship game. We describe the experimental setup in Section~\ref{sec:evaluation} and present the empirical evaluation in Section~\ref{sec:results}. Then, we expand the discussion of the algorithm's components in Section~\ref{sec:ablation}. Related work is presented in Section~\ref{sec:related} and we conclude in Section~\ref{sec:conclusions}.

%% file: preliminaries.tex
\section{Preliminaries}
\label{sec:preliminaries}
In this section we present the task of entity matching (Section~\ref{sec:EM}) and introduce the general framework of active learning (Section~\ref{sec:AL}). 

\begin{figure*}[t]
	\centering
	\includegraphics[width=0.86\textwidth]{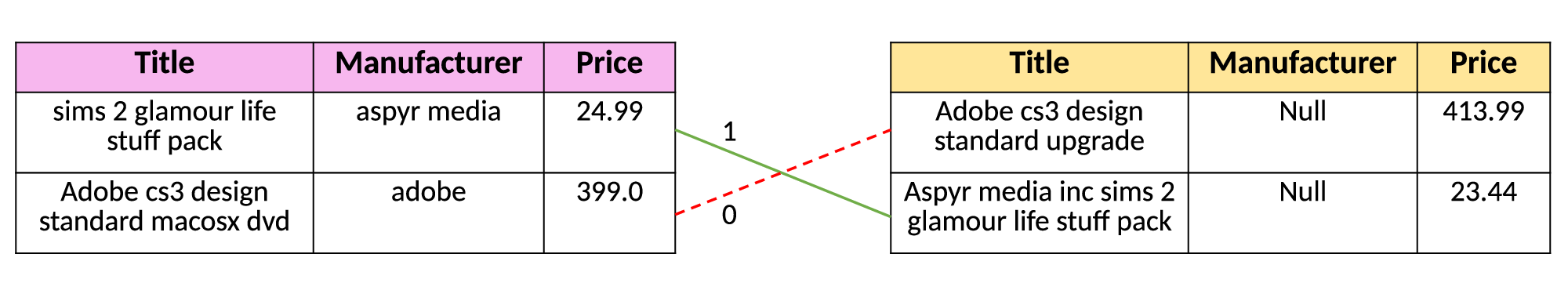}
		\vskip-.25in
	\caption{An example of entity matching between two tables. Candidate pairs are connected via a line where solid green (1) is a match and dotted red (0) a non-match.}
	\label{fig:EM_example}
\end{figure*}

\subsection{Entity Matching}
\label{sec:EM}
For convenience of presentation, and following~\cite{li2020deep}, we introduce entity matching as the task of matching entities between two clean datasets, such that in each dataset there are no entity duplicates (clean-clean matching). This definition can be easily extended to a single dirty dataset or multiple (clean or dirty) datasets. Consider two datasets $D_1$ and $D_2$ of entities (\emph{e.g.,} products, locations, academic papers, {\em etc}.). Given a pair of \tups $\left( r_1, r_2\right)  \in D_1 \times D_2$, the objective of entity matching is to determine whether $r_1 \in D_1$ and $r_2 \in D_2$ represent the same real-world entity. The essence of entity matching is to generate (train) a model that can accurately provide match/non-match predictions regarding \tup pairs.

\begin{example}
To illustrate, Figure~\ref{fig:EM_example} shows an instance taken from the Amazon-Google dataset. An example of a matching tuple pair (labeled with 1) involves the first \tup from the Amazon table (left) and the second \tup from the Google table (right). All other pairs, including the second \tup of the Amazon table and the first \tup of the Google table, are non-match pairs (labeled with 0).
\end{example}

Traditionally, the matching phase is preceded by a blocking phase~\cite{papadakis2020blocking}, which aims at reducing the computational effort of classifying the full set of possible pairs ($\left| D_1\right| \times \left|D_2\right|$). This is usually done by eliminating unlikely matches, keeping a relatively small set of candidate pairs. 
In this work we focus on the matching itself, assuming that the candidate pair set was already extracted using existing methods (see empirical comparison~\cite{papadakis2016comparative}). Each \tup is structured using a set of attributes $\left\lbrace A_i \right\rbrace$ 
and a \tup can be represented as a set of pairs $\left\lbrace (Att_i, Val_i)\right\rbrace $ for $1 \leq i \leq m$.
\begin{sloppypar}
In this work, we use transformer-based pre-trained language models, 
such as \emph{BERT}~\cite{devlin:2018bert} and its descendants (\emph{e.g.,} Roberta~\cite{liu:2019roberta} and DistilBERT~\cite{sanh2019distilbert}), which have shown state-of-the-art results in traditional natural language processing classification tasks and have been utilized for entity matching as well~\cite{li2020deep, brunner2020entity, miao2021rotom}. In such language models, a special $\left[ CLS\right]$ token is added at the beginning of the text, allowing the model to encode the input representation into a vector with dimensionality of 768. To yield a prediction, the embedding of the $\left[ CLS\right]$ token is pooled from the last transformer layer and injected as input into a fully connected network. In the context of entity matching, we follow~\cite{li2020deep} and serialize 
\tup pairs with a syntactic separation between them, where each \tup by itself is a serialization of its attribute-value pairs. 
\end{sloppypar}
\begin{example}
For example, the match candidate pair in Figure~\ref{fig:EM_example} is serialize as follows: 
\emph{"[CLS] [COL] title [VAL] sims 2 glamour life stuff pack [COL] manufacturer [VAL] aspyr media [COL] price [VAL] 24.99 [SEP] [COL] title [VAL] aspyr media inc sims 2 glamour life stuff pack [COL] manufacturer [VAL] [COL] price [VAL] 23.44"}. 
\end{example}

The model uses the extracted embeddings of the $\left[ CLS\right]$ token to make a final prediction regarding the pair. 
Since it serves as the connecting thread between the entire tokens sequence to its corresponding final prediction, the $\left[ CLS\right]$ token is widely treated as the {\em representative vector of the input}.    

\subsection{Active Learning}
\label{sec:AL}
\update{Supervised models, particularly those based on neural networks, require massive amounts of labeled data.} Active learning is widely employed \update{in} low resource settings to avoid the costly use of human \update{annotators}~\cite{meduri2020comprehensive, settles2009active}. Essentially, active learning focuses on iterative selection of samples to be labeled in a way that the selected ones are projected to provide significant informative power to the learning model. The selected samples are labeled by an oracle (\emph{e.g., a human annotator}) and then added to the training set.

A common selection criterion in active learning is the {\em certainty} a model assigns with samples~\cite{meduri2020comprehensive, kasai2019low, qian2017active}. Alongside the prediction itself, most classifiers also produce confidence values, where low confidence samples are believed to be more conductive to the model understanding. This principle was also employed in entity matching and entity resolution~\cite{kasai2019low, qian2017active, bogatu2021cost, jain2021deep}. For example, Kasai \emph{et al.}~\cite{kasai2019low} measure pair uncertainty using conditional entropy:
\begin{equation}
\label{eq:conditional_entropy}
H\left( p\right)=-plog\left( p\right) - \left(1-p\right)log\left( 1-p\right)
\end{equation}
where $p$ is the confidence \update{the model assigns} a given pair being a match. 
Another selection criterion in active learning is the {\em centrality} of samples, assuming more representative \update{data samples} are more informative to the model as well~\cite{zhang2021alg, zhu2008active}. Centrality can computed in multiple ways (\emph{e.g.,} betweenness centrality~\cite{freeman1977set}).

In this work, we use a graph structure to model a pair set, and apply both measures, accounting for vector space spatial considerations. For uncertainty, we extend Eq.~\ref{eq:conditional_entropy} to neighborhood agreement computation. As for centrality, we use \emph{PageRank}~\cite{page1999pagerank}, a well known measure that captures relative importance of a node in a graph by computing a probability distribution using \update{graph} topology.

%% file: model.tex
\begin{figure*}[t]
	\centering
	\includegraphics[width=0.755\textwidth]{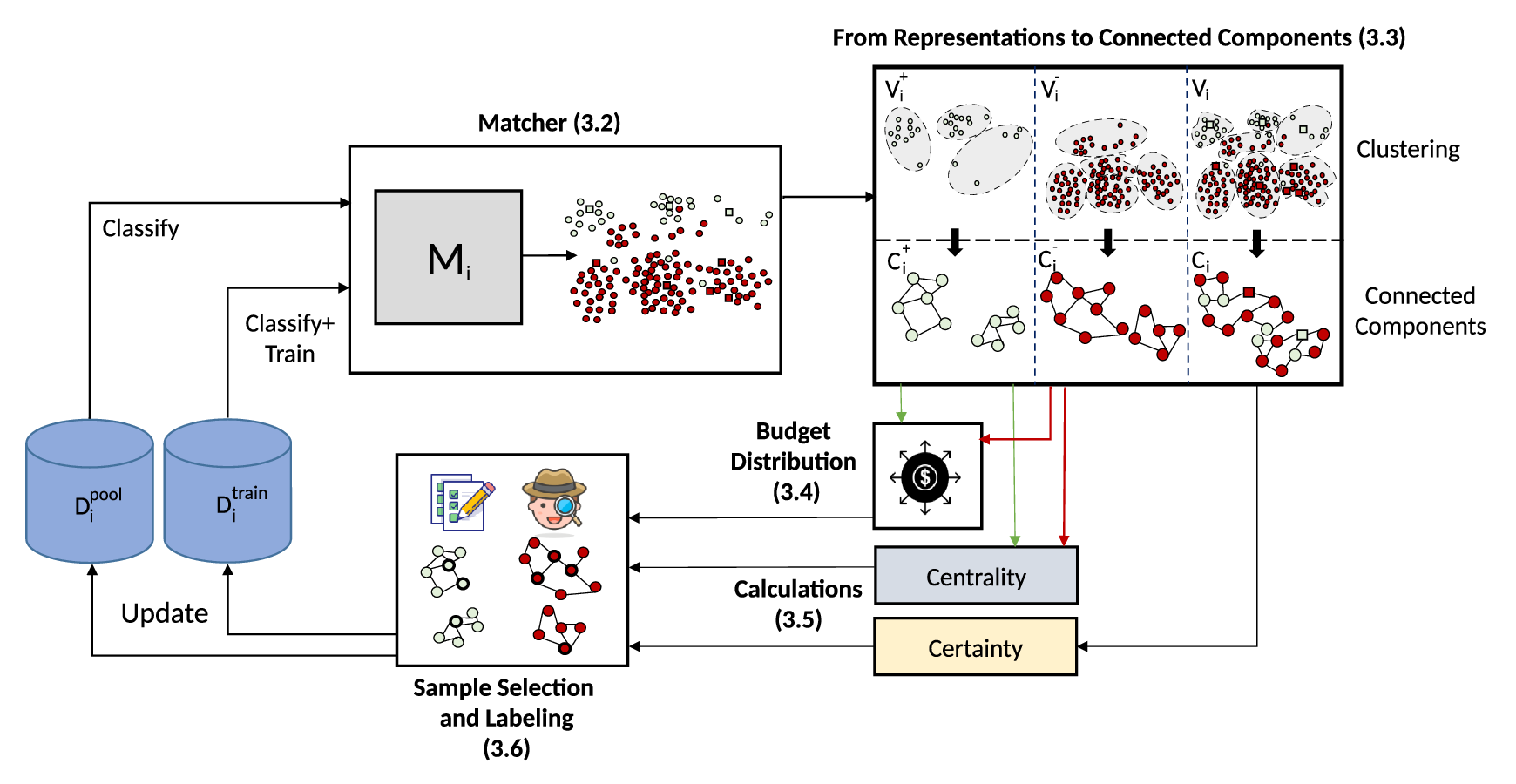}
	\vskip-.2in
	\caption{An Illustration of the battleship approach framework.}
	\vskip-.2in
	\label{fig:framework}
\end{figure*}

\section{Active Learning using the Battleship Approach}
\label{sec:model}


This section introduces an active learning algorithm to select, in an iterative manner, beneficial samples to label for solving entity matching problems.  
The proposed algorithm uses three selection criteria, namely {\em certainty}, {\em centrality}, and {\em correspondence}. The first two were introduced as general selection criteria for active learning in Section~\ref{sec:AL}. {\em Correspondence} handles a unique setting of data imbalance in entity matching, 
with a high percentage of no match pairs. Label imbalance interferes with the production of a quality model that generalizes well to new unseen data, especially for low resources. Our preliminary experiments show that if the quantity of matching pairs is insufficiently large the model often gets stuck in a local minimum where it only predicts no match labels. 
A crucial challenge of active learning in this domain is therefore to balance match and no match samples, so that the model can generalize well for both classes, overcoming this cold start challenge. 

\begin{table}[b]
	\begin{center}
		\scalebox{.78}{\begin{tabular}{cc}
				\midrule
				\textbf{Notation} & \textbf{Meaning} \\\midrule
				$B$  & Labeling budget per active learning iteration \\
				$I$ & Number of active learning iterations \\
				$i$ & Active learning iteration index \\
				$D$ & dataset \\
				$D_{i}^{train}$ & Labeled subset of $D$ in iteration $i$ \\
				$D_{i}^{pool}$ & Unlabeled subset of $D$ in iteration $i$ \\
				$M_i$ & Trained model in iteration $i$ \\
				$\mathbb{E}_i\left(j,p\right)$ & Pair representation of $\left( r_j, r_p \right)$ in iteration $i$\\
				$\phi\left( v \right)$ & Model confidence in the pair match\\
				$\pi\left( e \right)$ & The weight of edge $e$\\
				$G^+_i=\left( V^+_i,E^+_i\right)$ & Pair graph over match predicted samples from $D_{i}^{pool}$ \\
				$G^-_i=\left( V^-_i,E^-_i\right)$ & Pair graph over non-match predicted samples from $D_{i}^{pool}$ \\
				$G_i=\left( V,E_{i}\right)$ & Pair graph built upon $D$ \\
				$CC^*_i$ & Connected components of $G^*_i$ ($*\in \left\lbrace +,-\right\rbrace $) \\
				$CC_i$ & Connected components of $G_i$ \\	
				\hline
		\end{tabular}}
		\caption{Notation Table}
		\label{tab:symbols}
	\end{center}
\end{table}

The proposed sampling strategy uses a modified version of conditional entropy, expressing the induced uncertainty of a \tup pair with respect to spatial considerations, and PageRank centrality. To cope with the label imbalance we aim at pinpointing regions of the sampling space that are more likely to include match samples, in support of correspondence.
Then, we use the pooled sample representations to carefully select samples. From that point on, active learning plays a role in tuning the already existing model. Table~\ref{tab:symbols} summarizes the notations used throughout this work. 

\subsection{Overview}
\label{sec:overview}
A complete illustration of the proposed active learning algorithm is given in Figure~\ref{fig:framework}. We briefly outline the process first, and dive into details of the various components in the remainder of the section (see subsection numbering in Figure~\ref{fig:framework}).
Active learning algorithms are iterative, and we use $i\in\{0,1,\cdots,I\}$ to denote the iteration number. $D$, the initially unlabeled dataset of candidate pairs, is partitioned in each iteration $i$ into two disjoint pair sets, $D_{i}^{train}$ and $D_{i}^{pool}$, such that $D=D_{i}^{train} \cup D_{i}^{pool}$ and $D_{i}^{train} \cap D_{i}^{pool} = \emptyset$ (bottom left of Figure~\ref{fig:framework}). $D_{i}^{train}$ is an already labeled subset of $D$  while $D_{i}^{pool}$ represents the remaining unlabeled dataset of $D$.

In a single active learning iteration $B$ (denotes the labeling budget) new pairs are sent for labeling. 
Following previous works~\cite{kasai2019low, jain2021deep}, we assume the existence of labeled initialization seed $D_{0}^{train}$ (such that $D_{0}^{pool}=D \setminus D_{0}^{train}$), which is used for training the initial model $M_0$.
Samples from the target domain $D$ are first labeled with $M_0$ and then, through an iterative process, 
the labeled data in iteration $i$ ($D_{i}^{train}$) is used to train a model $M_{i}$ (mid top part of Figure~\ref{fig:framework}). The entire set, $D$, is then inserted into the new model, yielding a representative vector for each pair. Pairs are separated into connected components, built upon constrained K-Means clustering (right top part of Figure~\ref{fig:framework}). 
Each connected component receives a relative budget to its size and used for calculating pair certainty and centrality scores. The samples are carefully selected under budget limitations (mid bottom part of Figure~\ref{fig:framework}), considering the three selection criteria (certainty, centrality, and correspondence). 
New labeled data are then moved from $D_{i}^{pool}$ to $D_{i +1}^{train}$ and the iterative process repeats until the halting condition is met.

	
\subsection{Training a Matcher with Labeled Samples}
\label{sec:matcher}
At each iteration $i$ (including the initialization phase) the model that was trained with $D_{i-1}^{train}$ (top left part of Figure~\ref{fig:framework}) is now targeted to classify pairs from the entire set $D$. We inject all pairs $\left( r_i, r_j\right)  \in D$ through the model to extract $\mathbb{E}_i\left(j,p\right)$ and $\hat{y}_{j,p}$, the pair representation (the embeddings of the $\left[ CLS\right]$ token, taken from the last transformer layer) and the model's prediction, respectively. With each new active learning iteration, the model is trained with enriched training set, such that new labeled samples are added to previous ones. For example, after the first active learning iteration the model is trained with $D_{1}^{train}$, which consists of $D_{0}^{train}$ and $B$ new labeled samples, selected from $D_{0}^{pool}$. As the iterative process progresses, we expect $\mathbb{E}_i$ to contain a better pair representation, expressed in a clearer separation of match and no match samples.

\subsection{\Graphsnospace: From Spatial Representations to Connected Components}
\label{sec:from_satial_to_ccs}
The generation of pair representations is a cornerstone in selecting the most informative samples to be labeled. We hypothesize that a major share of informativeness for the entity matching task is a derivative of regional properties. Hence, we exploit the extracted representations in a way that allows us to pinpoint regional representatives. To do so, we build a graph derived from \tup pair vector-space representations and their similarities, and partition it into connected components as a tool of excavating spatial properties. We also make use of the graph to compute certainty and centrality scores (Sections~\ref{sec:Certainty} and~\ref{sec:Centrality}).

Let $G=\left( V,E\right)$ be a {\em \graphnospace} such that $V$ 
is a set of nodes representing \tup pairs. Each $v \in V$ is associated with its corresponding 
representative vector and a value $\nodeweight(v)$ 
that denotes the model confidence in the pair match. $E$ is a set of weighted edges, where existence of an edge $e=(u,v)\in E$ reflects spatial proximity of the pairs $u$ and $v$. The edge weight $\edgeweight(e)$ is the similarity score between $u$ and $v$, calculated using cosine similarity function.

\subsubsection{\textbf{Clustering Using Constrained K-Means}}
\label{sec:kmeans}
As a preparatory phase to the graph creation, we cluster the samples using K-Means over their representations. The motivation for this step is twofold. First, it partitions the vector-space into separate regions, which guarantees diverse sampling. Second, the edge creation process requires similarity comparisons between samples. Our algorithm allows comparisons only for samples that reside in the same cluster, hence significantly reducing the computational effort. While the second motivation also serves to motivate blocking in entity matching, it is the diversity that mainly drives the algorithm decision making.

We apply a constrained version of K-Means~\cite{bradley2000constrained} to avoid small clusters that cannot be represented under budget limitations, or alternatively, large clusters that demand multiple similarity comparisons. We set a minimal and maximal size for a cluster and select the $k$ value according to the Kneedle algorithm~\cite{satopaa2011finding} over the average sum of squared distance between the centroid of each cluster to its members. If the Kneedle algorithm fails to find a target value we select $k$ as the one that maximizes the silhouette score~\cite{thinsungnoena2015clustering}, a common clustering evaluation metric measuring intra-cluster cohesiveness comparing to inter-cluster separation.

\subsubsection{\textbf{Edge Creation}}
\label{sec:edge_creation}

The clustering of pair representations is employed for generating a representation graph. The structure of the graph is determined by the selection of hyperparameters defining node connectivity. The graph is utilized to capture the notions of certainty and centrality. Therefore, each node shall be connected to a minimal number of neighbors (for spatial-aware certainty calculation, as detailed in Section~\ref{sec:Certainty}) in a way that central nodes (with large number of adjacent representations in $\mathbb{E}_i$) are rewarded by being directly connected to additional nodes.

We connect each node (pair representation) to its $q$ (a hyperparameter of the model) nearest neighbors (in terms of cosine similarity score) among its cluster members. A large value of $q$ leads to a more robust certainty calculation and to a better graph connectivity, reducing the possibility of obtaining multiple small connected components that are under-represented under the budget distribution policy (Section~\ref{sec:budget_distribution}). However, it also increases the computational effort and might lead to a misrepresentation of cluster margins, which can harm the diversity. In addition, and for the sake of a better centrality estimation, we sort the remaining pairs within each cluster in descending order based on their similarity scores. We then generate edges for the closest \update{node pairs} in this sorted list. The ratio of selected node pairs is defined by a hyperparameter such that the total number of additional edges is proportional to the cluster size.
Intuitively, a more central node is more likely to be connected to a larger number of nodes. 
We assign an edge $e$ with its corresponding similarity score, denoted $\edgeweight(e)$, as its weight.
The resulting graph consists of multiple connected components, where each cluster yields one (or more) connected components.

\begin{figure}[t]
	\centering
	\includegraphics[width=0.755\columnwidth]{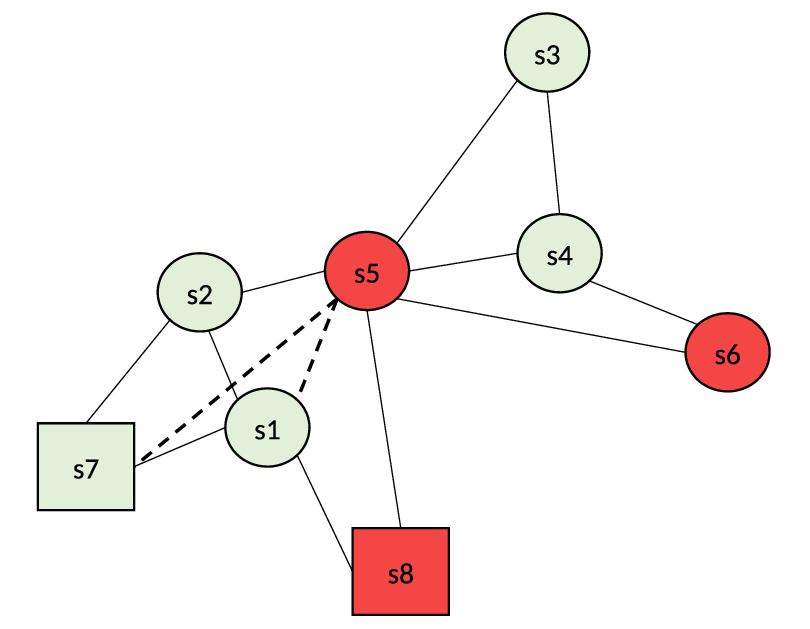}
	\vskip-.1in
	\caption{Edge creation and certainty calculation: Light green circles and squares represent match predicted and labeled samples, respectively. Red circles and squares represent no match predicted and labeled samples, respectively.}
	\vskip-.12in
	\label{fig:graph_creation}
\end{figure}

\begin{example}

Figure~\ref{fig:graph_creation} and Table~\ref{tab:pair_confidence_values} jointly offer an illustrative example of the edge creation process. Assume that $s1$-$s8$ are pair representations forming a single cluster, created according to the principles outlined in Section~\ref{sec:kmeans}. The samples $s1, s2, s3$ and $s4$ (light green circles) were predicted as match by the matcher (see Section~\ref{sec:matcher}), $s5$ and $s6$ (red circles) were predicted as no match and $s7$ and $s8$ were already labeled as match and no match, respectively, in a previous iteration. The values shown in Table~\ref{tab:pair_confidence_values} reflect the similarity scores between pair representations, while the values in the diagonal (blue cells) are the confidence scores assigned by the model (already labeled samples receive a confidence score of $1$), e.g., the similarity score between $s1$ and $s2$ is $0.9$ (relatively high as reflected by their closeness in Figure~\ref{fig:graph_creation}) and the matcher confidence of $s1$ being a match is $0.95$. Similarity score is symmetric, leading to a symmetric matrix. Assume that in this example $q=2$, namely each pair is automatically connected to its two nearest neighbors. In addition, assume that $0.15$ of the remaining potential edges are created (the closest ones among them).

\begin{table}[h]
	\begin{center}
		\scalebox{.9}{\begin{tabular}{|c|*{8}{c}|}
				\cline{1-9} & 
				\textbf{s1} & \textbf{s2} & \textbf{s3} & \textbf{s4} & \textbf{s5} & \textbf{s6} & \textbf{s7}  & \textbf{s8}   	\\ \cline{1-9} 
				\textbf{s1} & \cellcolor{blue!25}.95 & .9 & .5 & .6 & .85 & .5 & .9 & .82                 							\\ \cline{2-9} 
				\textbf{s2} &  & \cellcolor{blue!25}.92 & .55 & .58 & .92 & .45 & .83 & .6                             				\\ \cline{3-9} 
				\textbf{s3} &  &  & \cellcolor{blue!25}.96 & .75 & .67 & .56 & .4 & .38                         					\\ \cline{4-9}
				\textbf{s4} &  &  &  & \cellcolor{blue!25}.94 & .88 & .84 & .5 & .55                         						\\ \cline{5-9}
				\textbf{s5} &  &  &  &  & \cellcolor{blue!25}.98 & .57 & .63 & .65                         							\\ \cline{6-9}
				\textbf{s6} &  &  &  &  &  & \cellcolor{blue!25}.88 & .41 & .54                        								\\ \cline{7-9}
				\textbf{s7} &  &  &  &  &  &  & \cellcolor{blue!25}1 & .64                         									\\ \cline{8-9}
				\textbf{s8} &  &  &  &  &  &  &  & \cellcolor{blue!25}1                     										\\ \cline{9-9}
				\hline
		\end{tabular}}
	\end{center}
	\caption{Similarity scores for the samples presented in Figure~\ref{fig:graph_creation}. The values (blue cells) are the confidence scores assigned by the matcher for each sample.}
	\vskip-.12in
	\label{tab:pair_confidence_values}
\end{table}

Each pair is connected to at least two other pairs (its two nearest neighbors), denoted by solid lines. A pair can be connected to other pairs by more than two solid lines. For example, $s1$ is automatically connected to $s2$ and $s7$, since they have the highest scores in its similarity list (with $0.9$). $s1$ is also connected by a solid line to $s8$ since the former is among the two nearest neighbors of the latter (although $s8$ is only the fifth closest pair to $s1$). The total possible number of edges in this example is $\binom{8}{2}=28$, and $12$ edges are created in the first stage of connecting each pair to its two nearest neighbors. Therefore, $16$ possible edges are remaining, such that two ($\lfloor 0.15 \cdot 16 \rfloor=2$) additional edges shall be created. We rank those edges in a descending order, resulting in linking $s1$ to $s5$ ($0.85$) and $s5$ to $s7$ ($0.63$), denoted in the figure by dashed lines. It is noteworthy that we do not directly connect two labeled pairs, as they are not a target for the certainty calculations (Section~\ref{sec:Certainty}). Thus, $s7$ is not connected to $s8$ even though their similarity is higher than the similarity between $s5$ and $s7$.
\end{example}

\subsubsection{\textbf{Prediction and Heterogeneous Graphs}}
\label{sec:graphs}
To support effective sample selection that caters to the class imbalance problem, we suggest to separate match from no
	match samples. To do so, we create three graphs using the clustering mechanism described in Section~\ref{sec:kmeans}, each with a different set of nodes. 
In each iteration $i$, we partition $D_{i}^{train}$ into match labeled samples ($D_i^+$) and no match labeled samples ($D_i^-$). In a similar fashion, we also partition $ D_{i}^{pool}$ into $D^{pool+}_{i}=\{r,r^{\prime}\in D_{i}^{pool} \mid M_{i-1}(r,r^{\prime})=1\}$ and $ D_{i}^{pool-}=\{r,r^{\prime}\in  D_{i}^{pool} \mid M_{i-1}(r,r^{\prime})=0\}$, using the latest model to predict the labels of the unlabeled samples. Next, we create the three \graphs  $G^+_i=\left( V^+_i,E^+_i\right)$, where $V^+_i=D^{pool+}_{i}$ includes only the samples predicted to match, $G^-_i=\left( V^-_i,E^-_i\right)$, where $V^{pool-}_i=D^-_{i}$ contains samples predicted to be no match, and $G_i=\left( V,E_{i}\right)$, where $V=D$ contains all samples. Edges are generated, as detailed in Section~\ref{sec:edge_creation}. 

The first two graphs are prediction-based graphs, used for spatial sampling (Section~\ref{sec:budget_distribution}) and centrality calculations (Section~\ref{sec:Centrality}). The third is a heterogeneous graph, consisting of both labeled and unlabeled samples, disregarding their labels and predictions. It is used for certainty calculations (Section~\ref{sec:Certainty}).
We denote by $CC_i^{+}$, $CC_i^{-}$, and $CC_i$ the connected components sets that are generated from the three graphs and show how such a separation enhances the chances of successful sample selection.

\begin{example}
In the illustration shown in Figure~\ref{fig:framework} (mid top part of the figure) we use light green and red circles to denote match and no match predicted samples, respectively, and squares for already labeled samples (with the same color scheme). The squares play a role only in the heterogeneous graph (right top part of the figure), as only match and no match predicted samples appear in the graphs derived from $V^+_i$ and $V^-_i$, respectively.   
\end{example}


\subsection{Budget Distribution}
\label{sec:budget_distribution}
In a single active learning iteration, $B$ unlabeled pairs are selected to be labeled. We aim at a 
balanced pair selection, selecting both 
match and no match samples and sampling proportionally from dense and sparse spatial regions. Towards this end, we make use of the connected components that are created according to the description in Section~\ref{sec:from_satial_to_ccs} (see right bottom part of Figure~\ref{fig:framework}).

The match and no match predicted samples form separate connected component sets, $CC_i^{+}$ and $CC_i^{-}$.
Let $B$ be the \emph{labeling budget}, split into two distinct budgets, $B^{+}$ and $B^{-}$ ($B=B^{+}+B^{-}$), for expected match and expected no match pairs, respectively.
Intuitively, we would like to assign larger share of the budget to larger connected components. We define the budget share of a connected component $cc$ as follows:
\begin{equation}
	\label{eq:cc_budget_share}
	cc_{i,budget}= 
	\floor*{\frac{B^{*} \cdot\left| cc_{i}\right| }{\sum\limits_{cc^\prime_{i} \in CC_{i}^{*}}\left| cc^\prime_{i}\right| }}
\end{equation}
where the asterisk in $B^{*}$ and $CC_{i}^{*}$ stands for either $+$ or $-$, and $\left| cc_{i}\right|$ is the size of the connected component.
Since we round down the budgets (being a positive integer), 
any budget residue is randomly distributed among connected components.
\begin{example}
Assume that 3,000 samples were predicted as matching, split into 10 connected components, such that two of them consist of 500 samples each, four of 300 samples each and another four of 200 samples each. In addition, assume that $B^{+}=50$. According to the budget distribution policy of Eq.~\ref{eq:cc_budget_share} the first two (500 samples each) will be assigned with a budget of 8 ($ \floor*{\frac{50 \cdot 500}{3000}}$) each, the next four with a budget of 5 each ($ \floor*{\frac{50 \cdot300}{3000}} $) and the last four with a budget of 3 each ($ \floor*{\frac{50 \cdot 200}{3000}}$). The residue (2) is randomly allocated.
\end{example}

\subsection{Selection Criteria Calculations}
\label{sec:calculations}

Recall the three criteria for sample selection, namely certainty, centrality, and correspondence. Correspondence is supported by the generation of the graphs $G^+_i$ and $G^-_i$ (Section~\ref{sec:graphs}) and the balanced budget distribution between match and no match connected components (Section~\ref{sec:budget_distribution}). We focus next on certainty and centrality when selecting samples to be labeled (see bottom part of Figure~\ref{fig:framework}), adapting them to the unique needs of entity matching. 

\subsubsection{\textbf{Certainty}}
\label{sec:Certainty}
Conditional Entropy (Eq.~\ref{eq:conditional_entropy}) is one of the common methods for measuring certainty in active learning~\cite{jain2021deep, kasai2019low}. However, highly parameterized models such as transformer-based pre-trained language models tend to produce an uncalibrated confidence value~\cite{guo2017calibration, jiang2021can}, assigning mostly dichotomous values close to either 0 or 1. Such values provide little entropy differentiation, rendering pure measures such as conditional entropy unreliable for the selection mechanism. 
To overcome the dichotomous barrier, we propose to add a spatial interpretation to certainty by computing the disagreement of a pair with other pairs in its vicinity, using $G_i=\left( V,E_{i}\right)$ (Section~\ref{sec:graphs}) and its induced connected components $CC_i$ for an effective certainty computation. 

Let $G_{cc}=\left( V_{cc},E_{cc} \right)$ be a subgraph of $G_i$ where $V_{cc}$ is the set of nodes in $cc \in CC_i$, each representing a candidate pair, and $E_{cc}$ is a set of weighted edges labeled with the cosine similarity of the corresponding pair representations. $V^+_{cc}$ and $V^-_{cc}$ represent the pair set predicted to be, or already labeled as match and no match, respectively. Given a node $v \in cc \cap D_{i}^{pool}$, $N^{+}\left( v\right)$ denotes the set of nodes in $V^+_{cc}$ that are connected to $v$ and $N\left( v\right)$ denotes the entire set of nodes connected to $v$. The spatial confidence of a model $M_{i-1}$ regarding $v \in cc$ is the weighted average confidence of $v$'s neighbors with respect to its assigned prediction, defined as follows.
\begin{equation}
\label{eq:spatial_confidence}
\tilde{\phi}\left(v\right) = \frac{\sum\limits_{v^\prime \in N^{*}\left( v\right)}\pi \left(v, v^\prime\right) \cdot \phi \left(v^\prime \right)}{\sum\limits_{v^\prime \in N\left( v\right)}\pi \left(v, v^\prime\right) \cdot \phi \left(v^\prime \right)}
\end{equation}
where the asterisk is $+$ if the prediction is 1, otherwise it is $-$, $\pi \left(v, v^\prime\right)$ is the cosine similarity between $v$ and $v^\prime$ representations, and $\phi \left(v^\prime \right)$ represents the confidence in a label and is set to $1$ for all $v^\prime \in D_{i}^{train}$. For all other nodes, not in the training set, it is set to the confidence value assigned by $M_{i-1}$ to $v^\prime$. 

\begin{example}
Consider again Figure~\ref{fig:graph_creation} together with Table~\ref{tab:pair_confidence_values} and assume that we are interested in calculating $\tilde{\phi}\left(s1\right)$. $s1$ is predicted a match, hence the numerator consists of its match predicted or labeled neighbors ($s2$ and $s7$), while the denominator considers also the no match nodes ($s5$ and $s8$). By using the similarity and confidence scores of Table~\ref{tab:pair_confidence_values}, the desired score is $$\tilde{\phi}\left(s1 \right) = \frac{0.9 \cdot 0.92 + 0.9 \cdot 1}{0.9 \cdot 0.92 + 0.9 \cdot 1 + 0.85 \cdot 0.98 + 0.82 \cdot 1} = 0.51$$

\end{example}

Relying on our observation that regions in the latent space tend to be homogeneous, we employ Eq.~\ref{eq:conditional_entropy} and define the {\em spatial entropy} of a node $v \in cc$ as $H \left( \tilde{\phi} \right)$.
We combine both node's model-based prediction confidence and its spatial confidence, computing the final certainty score as a linear combination of the standard conditional entropy and the spatial entropy (both using Eq.~\ref{eq:conditional_entropy}), as follows.
\begin{equation}
\label{eq:uncertainty_score}
\mathbb{S}_{unc}\left(v\right) = \beta \cdot H\left( \phi \left(v \right) \right) + \left(1-\beta \right) \cdot H\left( \tilde{\phi}\left(v\right)\right)
\end{equation}
where $\beta$ is a weighting parameter ($0\leq \beta \leq 1$).

\subsubsection{\textbf{Centrality}}
\label{sec:Centrality}
We use PageRank ~\cite{page1999pagerank}, a well-known centrality measure for node's importance in a graph, originally used for Web retrieval. PageRank outputs a probability vector over the entire set of nodes, such that the higher the probability is, the more central is a node. Since edge directionality is important for PageRank, we produce two inversely directed edges for each edge in a connected components with the same edge weight (similarity score, see Section~\ref{sec:edge_creation}). 

Centrality is computed only over the available pool elements, $CC^+_i$ and $CC^-_i$, disregarding the labeled samples.
PageRank centrality of a node $v \in V_{cc}$ is calculated as follows.
\begin{equation}
\label{eq:PageRank}
\mathbb{S}_{cen}\left( v\right) = \rho \sum\limits_{v^{\prime} \in N\left( v\right)} A_{v, v^\prime}\frac{\mathbb{S}_{cen}\left( v^\prime\right)}{\sum\limits_{v^{\prime\prime} \in V_{cc}} A_{v^{\prime}, v^{\prime\prime}}} + \frac{1- \rho}{\left| V_{cc}\right|} 
\end{equation}
where $A$ is a weighted adjacency matrix and $\rho$ is a sampling parameter, traditionally used in PageRank to avoid dead-end situations~\cite{page1999pagerank}.

\subsection{\textbf{Sample Selection and Labeling}}
\label{sec:sample_selection}
To overcome possible scaling issues, we propose to rank pairs (nodes) in a descending scores order, according to both certainty and centrality. We define \sloppy $\Re_{cer}\left( v\right)$ and $\Re_{cen}\left( v\right)$ as the ranking of node $v$ according to its certainty and centrality score, respectively. Then, we weigh the overall ranking, as follows.
\begin{equation}
\label{eq:weighted_rank}
\alpha \cdot \Re_{unc}\left( v\right) + \left( 1- \alpha\right) \cdot \Re_{cen}\left( v\right)
\end{equation}
where $\alpha$ is a weighting parameter ($0\leq \alpha \leq 1$), of which we expand the discussion in Section~\ref{sec:alpha_analysis}.
For a connected component $cc$, the top $cc_{i,budget}$ pairs according to the weighted rank 
are selected to be labeled, removed from the $D_{i+1}^{pool}$, and inserted into $D_{i+1}^{train}$, ready to train the model for the next iteration. Similar to previous works~\cite{kasai2019low,jain2021deep, settles2009active, ren2021survey}, we assume the existence of a perfect labeling oracle, recognizing that in real-world settings a labeler might be exposed to biases that affect labeling accuracy~\cite{shraga2021learning}.

\subsection{Optimization with Weak Supervision}
\label{sec:weak_supervision}
In addition to the new labeled samples obtained in each iteration, we enrich the training set without exceeding the labeling budget $B$. To do so, we use
a weak supervision approach, where unlabeled samples are augmented into the training set with their corresponding model-based prediction, treated as a label.
By that, we allow the model to learn from a larger training set without using human labor for annotation.
\emph{Kasai et al.}~\cite{kasai2019low} implemented this approach by selecting the most confident pairs, namely those with the lowest conditional entropy values (Eq.~\ref{eq:conditional_entropy}). Following the principles presented in Section~\ref{sec:Certainty}, we define the most confident pairs in a spatial-aware fashion, namely the selected samples are those that minimize the value of Eq.~\ref{eq:uncertainty_score}. To enhance diversity of sampling, the label-wise budget ($\frac{B}{2}$ for each of match and no match predicted sample sets) is distributed over the connected component (Section~\ref{sec:from_satial_to_ccs}) using the same procedure in Section~\ref{sec:budget_distribution}. As this addition was proven to be effective in our preliminary experiments, we have incorporated it into the entire set of experiments.

%% file: evaluation.tex
\section{Experimental Setup}
\label{sec:evaluation}
In this section we detail the benchmarks and evaluation methodology used to asses the performance of the battleship approach.

\subsection{Datasets}
\label{sec:datasets}
We use six publicly available datasets, from different domains. To be consistent with recent works~\cite{konda2016magellan, li2020deep}, we assume that a set of candidate \tup pairs is given, possibly a result of a blocking phase. A summary of the datasets (referring to the complete training set sizes) is given in Table~\ref{tab:Datasets} and detailed next.



\noindent \textbf{Walmart-Amazon and Amazon-Google:} The datasets of Walmart-Amazon and Amazon-Google, taken from the Magellan data repository~\cite{konda2016magellan},\footnote{\url{https://github.com/anhaidgroup/deepmatcher/blob/master/Datasets.md}} are well-known for evaluation of entity matching solutions.
We use the training, validation and test sets, with the ratio of 3:1:1 provided by the benchmark used in previous works (\emph{e.g.,}~\cite{mudgal2018deep,li2020deep}). 
The relative part of matching pairs (see Table~\ref{tab:Datasets}) is fairly small, turning the initialization of the model into a challenging task.

\noindent \textbf{WDC Cameras and WDC Shoes:} The web data commons (WDC) dataset~\cite{primpeli2019wdc} contains product data extracted from multiple e-shops\footnote{\url{http://webdatacommons.org/largescaleproductcorpus/v2/index.html}\label{fn:wdc}} and split into four categories. Following~\cite{li2020deep}, we use only the product title, ignoring other attributes. We focus on the medium size datasets and the product categories of Shoes and Cameras. We keep the exact same partition as~\cite{li2020deep}, with a test set of $\sim 1,100$ pairs per category dataset, while the remaining pairs are split into training and validation with the ratio of 4:1. 

\noindent \textbf{ABT-Buy:} 
ABT-Buy is also a product dataset. Unlike the rest of the experimented datasets, which are all structured, ABT-Buy contains long textual inputs. The task is to match company homepages to Wikipedia pages describing companies. We use the same 3:1:1 partiotion of the data, as used by~\cite{mudgal2018deep}. 

\noindent \textbf{DBLP-scholar:} 
DBLP-scholar is also a widely used benchmark for entity matching, containing bibliographic data from multiple sources. Whereas DBLP and is a high quality dataset, Google scholar is constructed using automatic Web crawling, hence lacking a rigorous data cleaning process. this benchmark also comes with a predefined training, validation and test sets, with the ratio of 3:1:1~\cite{konda2016magellan}.

\begin{table}[htpb]
	\begin{center}
		\scalebox{.995}{\begin{tabular}{|c|ccc|}
				\hline
				\textbf{Dataset} & \textbf{Size} &  \textbf{$\%$Pos} & \textbf{\#Atts}\\\hline
				 Walmart-Amazon & 6,144 & 9.4\% & 5\\\hline
				 Amazon-Google & 6,874 & 10.2\% & 3\\\hline
				 Cameras & 4,081 & 21.0\% & 1\\\hline
				 Shoes & 4,505 & 20.9\% & 1\\\hline
				 ABT-Buy & 5,743 & 10.7\% & 3 \\\hline
				 DBLP-Scholar & 17,223 & 18.6\% & 4\\\hline
			
		\end{tabular}}
	\end{center}
	\caption{Statistics of the datasets used in our experiments. The size values refer to the training sets.}
	\label{tab:Datasets}
\end{table}	

\subsection{Implementation Details}
\label{sec:setup}
Experiments were performed on a server with 2 Nvidia Quadro RTX 6000 and a CentOS 7 operating system. Our implementation is available in a git repository.\footnote{\url{https://github.com/BarGenossar/The-Battleship-Approach-to-AL-of-EM-Problem}}

\noindent \textbf{Model and Optimization:} We adopted the publicly available\footnote{\url{https://github.com/megagonlabs/ditto}} implementation of DITTO~\cite{li2020deep}, and modified it to yield, alongside predictions, also pair representations and confidence values. In each active learning iteration $i\in\{0,1,\cdots,I\}$ we train a model with the updated training set $D_{i}^{train}$ (see Section~\ref{sec:matcher}). 
For the first five datasets (Table~\ref{tab:Datasets}) we set the number of epochs to 12, while for the sixth (DBLP-Scholar) we run only over 8 epochs. These number were selected after preliminary experiments showing no significant improvement (if any) is achieved by using a larger number of epochs. 
The parameters of DITTO in an active learning iteration are initialized without using the values of previous iterations, and set according to the best F1 score achieved on the validation set. We use a maximum input length of 512 tokens, the maximal possible length for BERT-based models~\cite{devlin:2018bert}, and a batch size of 12. The model is trained with \emph{AdamW} optimizer~\cite{loshchilov2017decoupled} with a learning rate of $3e-5$.

The focus of this work is on the active learning's selection mechanism rather than the neural network architecture. Therefore, we use the basic form of DITTO, without optimizations~\cite{li2020deep}, fine-tuning a pre-trained language model (RoBERTa~\cite{liu:2019roberta}) on the specific task. For each configuration we report the average F1 values, calculated over 3 different seeds.

\noindent \textbf{Active Learning and Pair Selection:} We run 8 active learning iterations for all datasets, where $B$, the pair labeling budget per iteration, is fixed at 100, as well as the weak labels budget (see Section~\ref{sec:weak_supervision}). \update{Similar to previous works~\cite{kasai2019low, jain2021deep}, we start with labeled initialization seed $D_{0}^{train}$,  consists of $\frac{\left| B\right|}{2}$ (in our case $50$) samples for both matching pairs and non-matching pairs.} Following the principles presented in Section~\ref{sec:model}, we aimed at equipping the model with a balanced set of pairs. Since match labels are harder to discover, especially in the initial active learning iterations (Section~\ref{sec:intro}), we set the positive budget $B^{+}$ as $B \cdot max \left\lbrace 0.8-\frac{1}{20} \cdot i, 0.5\right\rbrace $, where $i$ is the active learning iteration number. By that, we increase the chances of feeding the model with relatively large share of positive samples, contributing to its generalizability, especially in the early iterations.

A pair is represented with a 768 dimensional vector, pooled of the last hidden layer of the model. 
The size of a cluster (Section~\ref{sec:kmeans}) ranges from $0.05$ to $0.15$ of the number of samples against which the graph is created ($\left| V^+_i\right| $, $\left| V^-_i\right| $ or $\left| V\right| $). 
The nearest neighbors calculations (Section ~\ref{sec:edge_creation}) are implemented with FAISS library~\cite{johnson2019billion}. The graphs are created such that each node is connected to its 15 nearest neighbors. In addition, the top $3\%$ of the rest of the sample pairs are also connected (Section~\ref{sec:edge_creation}).

\begin{figure*}[htpb]
	\centering
	\begin{subfigure}[b]{0.33\textwidth}
		\centering
		\includegraphics[width=\textwidth]{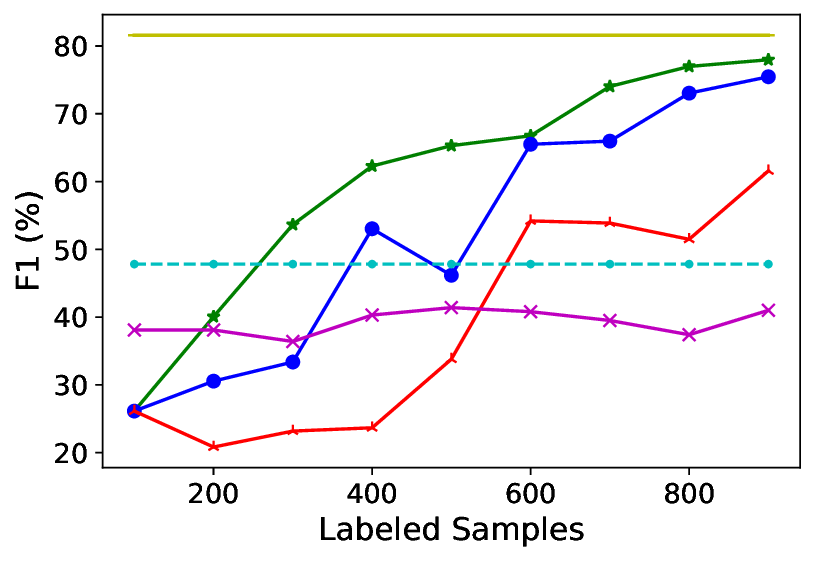}
		\caption{Walmart-Amazon}
		\label{fig:Walmart-Amazon_F1}
	\end{subfigure}
	\hfill
	\begin{subfigure}[b]{0.33\textwidth}
		\centering
		\includegraphics[width=\textwidth]{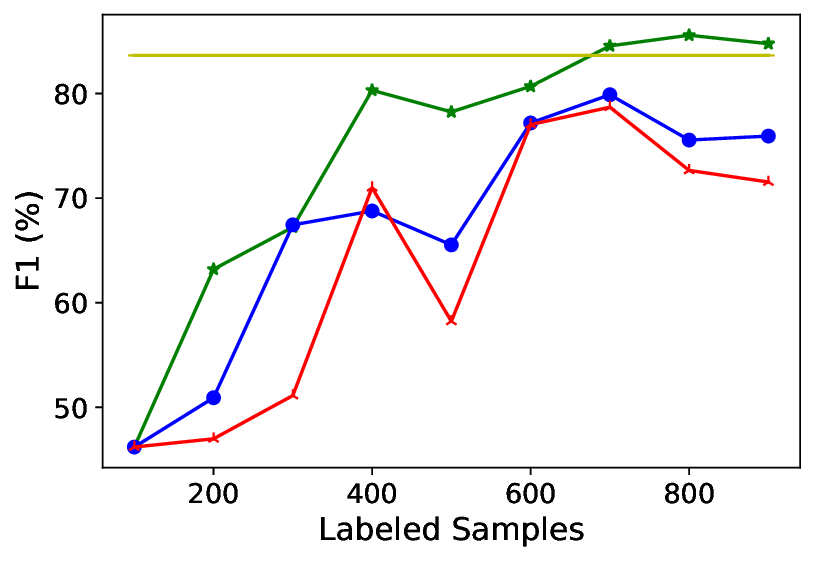}
		\caption{WDC Cameras}
		\label{fig:cameras_F1}
	\end{subfigure}
	\hfill
	\begin{subfigure}[b]{0.33\textwidth}
		\centering
		\includegraphics[width=\textwidth]{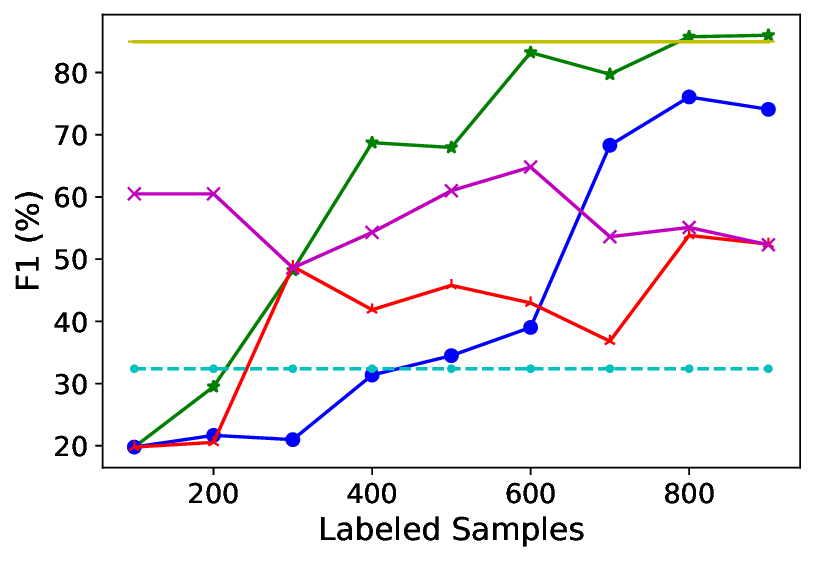}
		\caption{ABT-Buy}
		\label{fig:DBLP-ACM_F1}
	\end{subfigure}
	\begin{subfigure}[b]{0.33\textwidth}
		\centering
		\includegraphics[width=\textwidth]{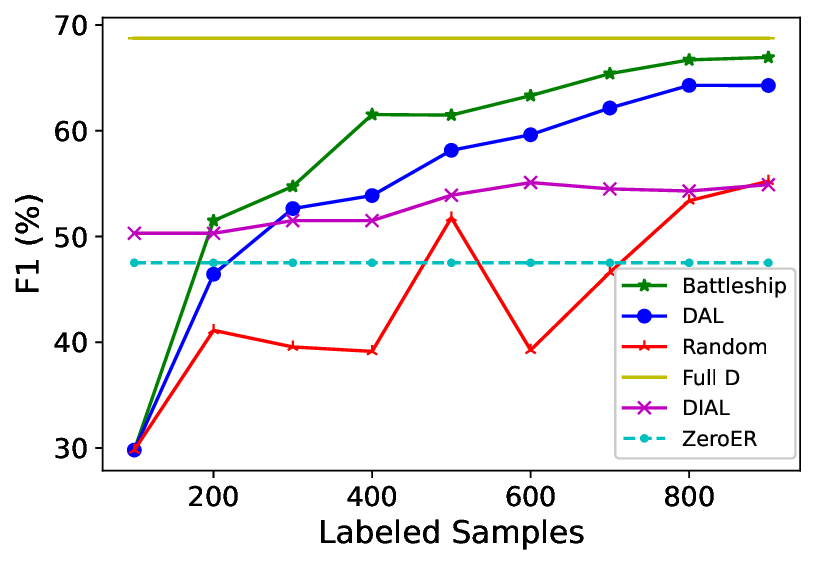}
		\caption{Amazon-Google}
		\label{fig:Amazon-Google_F1}
	\end{subfigure}
	\hfill
	\begin{subfigure}[b]{0.33\textwidth}
		\centering
		\includegraphics[width=\textwidth]{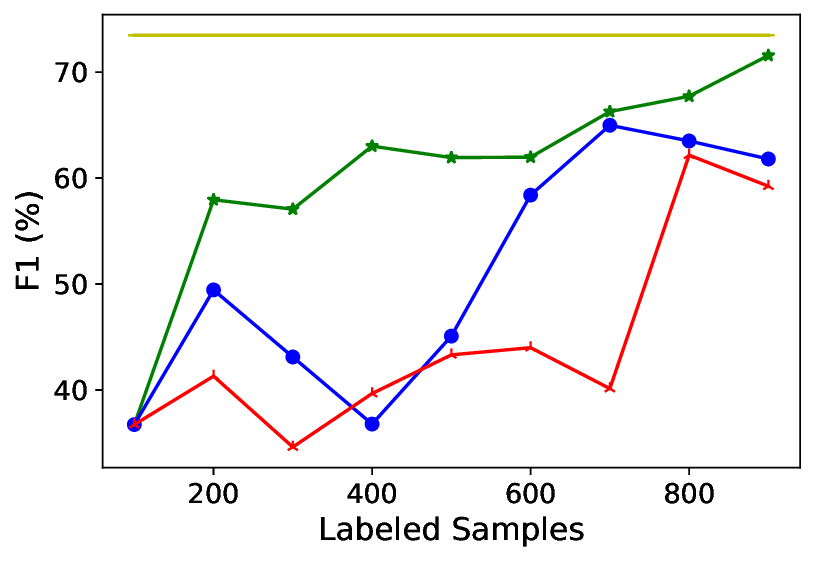}
		\caption{WDC Shoes}
		\label{fig:shoes_F1}
	\end{subfigure}
	\hfill
	\begin{subfigure}[b]{0.33\textwidth}
		\centering
		\includegraphics[width=\textwidth]{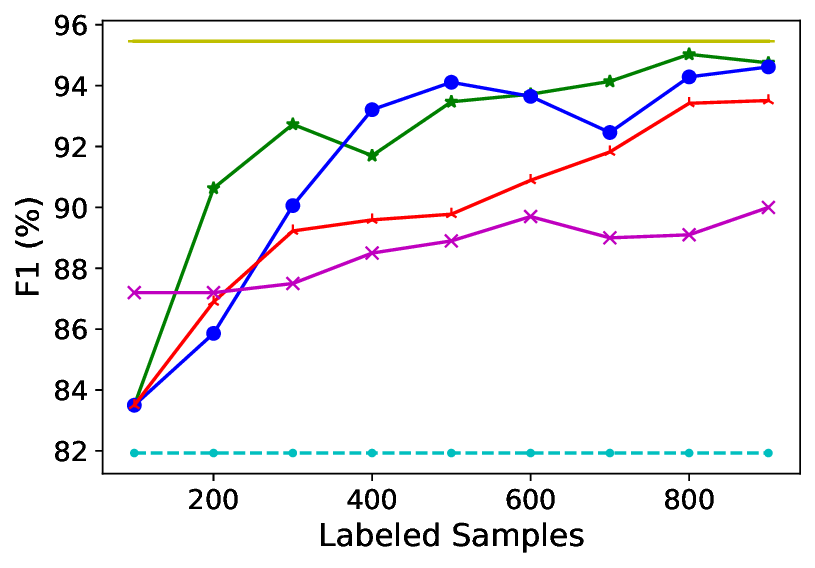}
		\caption{DBLP-Scholar}
		\label{fig:DBLP-GoogleScholar_F1}
	\end{subfigure}
	\caption{Performance in terms of F1 (\%) vs cumulative number of labeled samples. The Full D line represents the performance with completely available training data (the datasets sizes are displayed in Table~\ref{tab:Datasets}}
	\label{fig:F1vslabeled}
\end{figure*}


\subsection{Baselines Methods}
\label{sec:baseline_methods}
We compare the battleship approach with several baseline methods to active learning of the entity matching problem. We report the results over the same test set for all approaches.

\begin{itemize}
	\item \textbf{Random:} 
	A na\"ive baseline where samples are randomly drawn from the available pool, considering neither the predictions of the model nor the benefits of pair representations.
	\item \textbf{DAL} (Deep Active Learning)~\cite{kasai2019low}: In each active learning iteration, $\frac{B}{2}$ no match predictions and $\frac{B}{2}$ match predictions are labeled. Selected samples are the most uncertain (those maximizing the value of Eq.~\ref{eq:conditional_entropy}). In addition, DAL uses weak-supervision mechanism, augmenting the training set with $\frac{k}{2}$ match and no match high-confidence samples, with their assigned prediction (see Section~\ref{sec:weak_supervision}). In the absence of implementation, we implemented a version of DAL according the to guidelines presented in~\cite{kasai2019low}, without the adversarial transfer learning component since source domain data is not available in our settings. To align DAL with our proposed approach, we train the model as described in Section~\ref{sec:matcher}, using DITTO~\cite{li2020deep} as the neural network matcher.
	\item \textbf{DIAL} (Deep Indexed Active Learning)~\cite{jain2021deep}: DIAL offers an approach of co-learning embeddings to maximize recall (for blocking) and accuracy (for matching) using index-by-committee framework. Pre-trained transformer language models are used to create \tup representations, employed for vector-space similarity search to produce candidate pairs set. Then, pair selection is performed using uncertainty sampling. This characteristic differentiate DIAL from other approaches, as they are given a fixed set of candidate pairs. We test DIAL with the published implementation,\footnote{\url{https://github.com/ArjitJ/DIAL}} starting with 128 labeled samples, equally divided into matches and no matches. In our experiments, we do not evaluate DIAL on the WDC product datasets, which do not use a structure of two tables, a required input for the blocker.

\end{itemize}
In addition, we compare our approach to two methods that represent the extremes in terms of labeling resources, with the purpose of showing 
our approach to be competitive over the whole spectrum of resource possibilities.

\begin{itemize}
	\item \textbf{ZeroER} (Entity Resolution using Zero Labeled Examples)~\cite{wu2020zeroer}: ZeroER is an unsupervised approach that 
	relies on the assumption that similarity vector for match pairs should differ from that of no match pairs. A core difference between their work and ours is that they build model-agnostic feature vector to capture \tup pairs similarity, while we extract the representation yielded by a limited resource trained model. As the WDC datasets do not fit into the required input type of ZeroER, we do not evaluate this method on these datasets.
	\item \textbf{Full D:} We train DITTO over the complete training set \update{(see Table~\ref{tab:Datasets})}, assuming no lack of resources. 
\end{itemize}

%% file: experiments_results.tex
\section{Empirical Evaluation}
\label{sec:results}
We provide a thorough empirical analysis of the battleship approach. We start by comparing its performance to the multiple baselines (Section~\ref{sec:prog_f1}) and continue with running time report (Section~\ref{sec:running_time}).

\subsection{Battleship and Baselines Performance}
\label{sec:prog_f1}
Figure~\ref{fig:F1vslabeled} shows the F1 score with respect to the number of labeled samples for all datasets.
The results reported here for the battleship approach are the average performance for the model with three values of $\alpha$ ($0.25. 0.5, 0.75$), which determines the relative weight between certainty and centrality for the overall ranking (Eq.~\ref{eq:weighted_rank}). For each of these $\alpha$ values we fixed $\beta$, which determines the relative weight between local and spatial entropy in the certainty score (of which we expand the discussion in Section~\ref{sec:beta_analysis}), at $0.5$ . 

For Walmart-Amazon and Amazon-Google, the battleship approach outperforms other active learning baselines, performing almost as good as the Full D model, albeit with less labeled samples.
The battleship approach beats its best baseline (DAL) by a margin of $3.3\%$ for Walmart-Amazon and $4.1\%$ for Amazon-Google at the end of the active learning process (900 labeled samples).
For both datasets the performance of the battleship approach significantly increases during the first iterations, with a decreasing improvement rate later. The battleship approach focuses on spotting match pairs, particularly in the early stages of the iterative active learning process. Since the number of match pairs in both datasets is limited (see Table~\ref{tab:Datasets}), the battleship approach consumes them early on, adding more no match pairs as the process progresses, which undermines the principle of a balanced pool. 
 
 \begin{table}[t]
 	\centering
 	\scalebox{.62}{\begin{tabular}{|c|cccccccc|}
 			\hline
 			\textbf{Type} &\textbf{Model} & \textbf{$\#$Labels} & \textbf{Walmart} & \textbf{Amazon}   &   \textbf{WDC} & \textbf{WDC} & \textbf{ABT-} & \textbf{DBLP} \\
 			& & &\textbf{Amazon} & \textbf{Google}   & 	\textbf{Cameras}		  & \textbf{Shoes}	  & \textbf{Buy} & \textbf{Scholar} \\\hline
 			\multirow{1}{*}{\begin{tabular}[c]{@{}c@{}}Unsupervised\end{tabular}} 
 			&\multirow{1}{*}{\begin{tabular}[c]{@{}c@{}} ZeroER\cite{wu2020zeroer} \end{tabular}}
 			& 0 & 47.82 & 47.51 & - & - & 32.39 & 81.93 \\\cline{2-9}
 			\hline
 			
 			\multirow{1}{*}{\begin{tabular}[c]{@{}c@{}}Supervised\end{tabular}} 
 			&\multirow{1}{*}{\begin{tabular}[c]{@{}c@{}} Full D \end{tabular}}
 			& $\left| D\right| $ & 81.60 & 68.75 & 83.65 & 73.48 & 84.95 & 95.46 \\\cline{2-9}
 			\hline
 			
 			\multirow{8}{*}{\begin{tabular}[c]{@{}c@{}}Active \\ Learning\end{tabular}} 
 			&\multirow{2}{*}{\begin{tabular}[c]{@{}c@{}} Random \end{tabular}}
 			& 500 & 33.79 & 51.77& 58.22 & 43.31 & 45.79 & 89.78 \\
 			&& 900 & 61.57 & 55.23& 71.54 & 59.23 & 52.42 & 93.51 \\\cdashline{2-9}
 			&\multirow{2}{*}{\begin{tabular}[c]{@{}c@{}}DAL\cite{kasai2019low}\end{tabular}} 
 			& 500 & 46.17 & 58.15& 65.53 & 45.08 & 34.49 & \textbf{94.11} \\
 			&& 900 & 75.47 & 64.28& 75.93 & 61.80 & 74.08 & 94.62 \\\cdashline{2-9}
 			&\multirow{2}{*}{\begin{tabular}[c]{@{}c@{}}DIAL\cite{jain2021deep}\end{tabular}} 
 			& 500 & 41.40 & 53.90 & - & - & 61.30 & 88.90 \\
 			&& 900 & 41.00 & 54.90 & - & - & 52.30 & 90.00 \\\cline{2-9}
 			&\multirow{2}{*}{\begin{tabular}[c]{@{}c@{}}\cellcolor{blue!25}Battleship\end{tabular}} 
 			& 500 & \textbf{65.30} & \textbf{61.48} & \textbf{78.24} & \textbf{61.93} & \textbf{67.95} & 93.47 \\
 			&& 900 & \textbf{77.98} & \textbf{66.94} & \textbf{84.76} & \textbf{71.57} & \textbf{85.99} & \textbf{94.75} \\\cline{2-9}
 			\hline
 	\end{tabular}}
 	\caption{F1 values for varying labeled samples set size.}
 	\label{tab:F1_TAB}
 \end{table}

The battleship approach outperforms the baselines over WDC (both cameras and shoes) as well. Furthermore, it also outperforms the Full D model ($84.76$ {\em vs.} $83.65$) for the cameras dataset. A possible explanation to the success of the battleship approach over these two datasets might be their relative high ratio of matching pairs, which helps the selection mechanism to obtain balanced sampling throughout the training phase.
For ABT-buy, the battleship approach also surpasses the Full D model ($85.99$ {\em vs.} $84.95$), in addition to the baselines ($74.08$ for DAL). 
DBLP-Scholar is the only dataset of which the battleship approach \update{is almost tied with} the best baseline (DAL) ($94.75$ {\em vs.} $94.62$), both trail by a small margin behind the Full D model ($95.46$). 

For all datasets, the battleship approach improves upon the baseline early on, which we attribute to balance sampling that enables overcoming the cold-start problem for low-resource entity matching tasks. It can also be seen the the battleship approach requires at most two iterations to surpass the unsupervised approach of ZeroER.
We note that the reported results for DITTO~\cite{li2020deep} (here as the Full D model), DIAL~\cite{jain2021deep} and ZeroER~\cite{wu2020zeroer} differ from the ones reported in the respective papers. For the former, we used the publicly available code without optimizations. For the second, we used the code provided by the authors\footnote{\url{https://github.com/ArjitJ/DIAL}} and followed the instructions in the paper to reproduce it in our setting. As for the latter, we report the results we obtained by using publicly available code\footnote{\url{https://github.com/ZJU-DAILY/ZeroMatcher}} only over the test set, whereas the those reported in the paper~\cite{wu2020zeroer} refer to the training, validation and test sets combined. 



Table~\ref{tab:F1_TAB} emphasizes the effectiveness of the battleship approach under low resource limitation. In this table it is compared, alongside active learning baselines, with the fully trained model and with unsupervised approach, namely ZeroER. The table shows F1 performance with 500 (after four active learning iterations out of eight) and 900 (last active learning iteration) labeled samples for active learning methods.
Except for DBLP-Scholar, the battleship approach beats the active learning baselines with both 500 and 900 labeled samples, while for the second it also beats the fully trained model over WDC Cameras and ABT-Buy. ZeroER, although does not rely on labeled samples at all, performs better than the baselines with 500 samples over the Walmart-Amazon dataset.

In Table~\ref{tab:AUC_TAB} we estimate the performance of the battleship approach over the learning course using the Area Under Curve measurement~\cite{baram2004online, shraga2021limited}, calculated against the F1 plot. We notice, again, the significant gap between the battleship approach and the active learning baselines, as it is the most dominant method over all dataset. \update{It is also noteworthy that the battleship approach beats its best competitor (DAL) over the DBLP-Scholar, despite it is the only dataset in which the second outperforms the first in terms of F1 with 500 samples. The reason for that is the preferable performance of the battleship approach against DAL during the first iterations.}

\begin{table}[htpb]
	\centering
	\vskip 0.15in
	\scalebox{.83}{\begin{tabular}{|c|cccccc|}
			\hline
			\textbf{Model} & \textbf{Walmart} & \textbf{Amazon}   &   \textbf{WDC} & \textbf{WDC} & \textbf{ABT-} & \textbf{DBLP} \\
			& \textbf{Amazon} & \textbf{Google}   & 	\textbf{Cameras}		  & \textbf{Shoes}	  & \textbf{Buy} & \textbf{Scholar} \\\hline
			 Random & 304.86 & 353.32 & 514.56 & 353.14 & 326.73 & 720.13 \\\hline
			 DAL & 418.46 & 444.19 & 546.33 & 410.55 & 338.88& 732.70 \\\hline
			 DIAL & 313.45 & 423.70 & - & - & 454.30 & 708.50 \\\hline
			 \cellcolor{blue!25}Battleship & \textbf{491.15} & \textbf{473.03} & \textbf{605.25} & \textbf{490.06} & \textbf{515.96} & \textbf{740.54} \\\hline
			
	\end{tabular}}
	\caption{AUC for the F1 plots.}
	\label{tab:AUC_TAB}
\end{table}

\subsection{Runtime Analysis}
\label{sec:running_time}
\begin{figure}[htpb]
	\centering
	\includegraphics[width=.999\columnwidth]{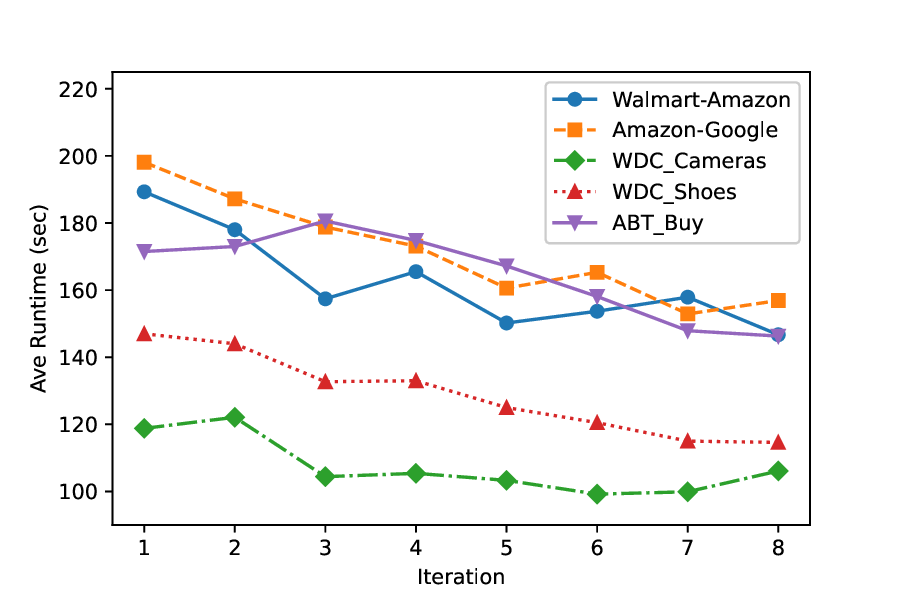}
	\caption{Runtime of the battleship approach {\em vs.} iteration.}
	\label{fig:runtime_plot}
\end{figure}

\update{The graph in Figure~\ref{fig:runtime_plot} illustrates the average running time (seconds) for the battleship approach with a fixed $\beta$ value (.5) and varying $\alpha$ values in the set $\left\lbrace 0.25, 0.5, 0.75\right\rbrace $, across different datasets, as a function of the iteration number. DBLP-Scholar was excluded 
due to its significant variation in scale, ranging from 430 to 549 seconds per iteration, that obscures the clarity of the figure. As observed
, the runtimes decrease as the active learning process progresses. This is attributed to the major impact of K-Means clustering on the overall runtime. As the active learning process continues, the pool of available data for generating the predicted match and non-match graphs (as discussed in Section~\ref{sec:graphs}) becomes smaller. Consequently, the K-Means computations are performed on reduced pair set, leading to improved efficiency.}
 
\update{In our analysis, we observed that DAL exhibits considerably faster performance, taking only a few seconds per iteration. This efficiency can be attributed to the absence of spatial considerations in this method, as the K-Means clustering step consumes the majority of the running time in the battleship approach. While there are methods such as LSH~\cite{gionis1999similarity} and HNSW~\cite{malkov2018efficient} that can reduce the computational effort of K-Means through approximate nearest neighbor searches, we focuse our attention on other aspects in this study and aim to explore this aspect in future work.}

\update{Additionally, we witnessed in the preliminary experiments that the number of nearest neighbors per node and the remaining pairs ratio (Section ~\ref{sec:edge_creation}) also influence the running time. Reducing these parameters may lead to a tradeoff since they directly impact the centrality and certainty computations in the battleship approach. While optimizing the running time is essential in real-world applications, we need to balance it with the effectiveness of the algorithm's core components to achieve accurate results.}

%% file: components_analysis.tex
\section{Components Analysis}
\label{sec:ablation}
In this section we focus on analyzing the various design choices described in Section~\ref{sec:model}, and how they impact the overall algorithm performance. Section~\ref{sec:beta_analysis} focuses on the local {\em vs.} spatial certainty tradeoff. The contribution of centrality and certainty is analyzed in Section~\ref{sec:alpha_analysis} and the correspondence effect is elaborated upon in Section~\ref{sec:correspondence_effect}. Section~\ref{sec:ws_analysis} delves into the impact of the weak supervision component. Then, we conclude the discussed tradeoffs in Section~\ref{sec:ablation_conclusion}.

\subsection{Local {\em vs.} Spatial Certainty}
\label{sec:beta_analysis}
To assess the impact of local and spatial certainty, we ran experiments with $\beta \in \left\lbrace 0, 0.5, 1\right\rbrace$ (see Eq.~\ref{eq:uncertainty_score}). Due to space limitations we present here only the results for Walmart-Amazon and Amazon-Google.
For all three values of $\beta$ we fixed $\alpha$ at $0.5$. 

\begin{figure}[htpb]
	\centering
	\begin{subfigure}[b]{0.73\columnwidth}
		\centering
		\includegraphics[width=\columnwidth]{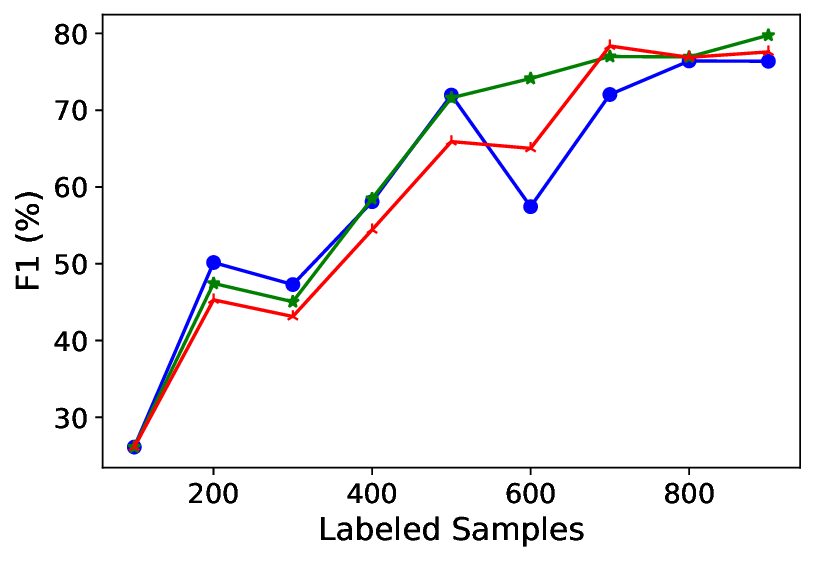}
		\caption{Walmart-Amazon}
		\label{fig:Walmart-Amazon_beta}
	\end{subfigure}
	\hfill
	\begin{subfigure}[b]{0.73\columnwidth}
		\centering
		\includegraphics[width=\columnwidth]{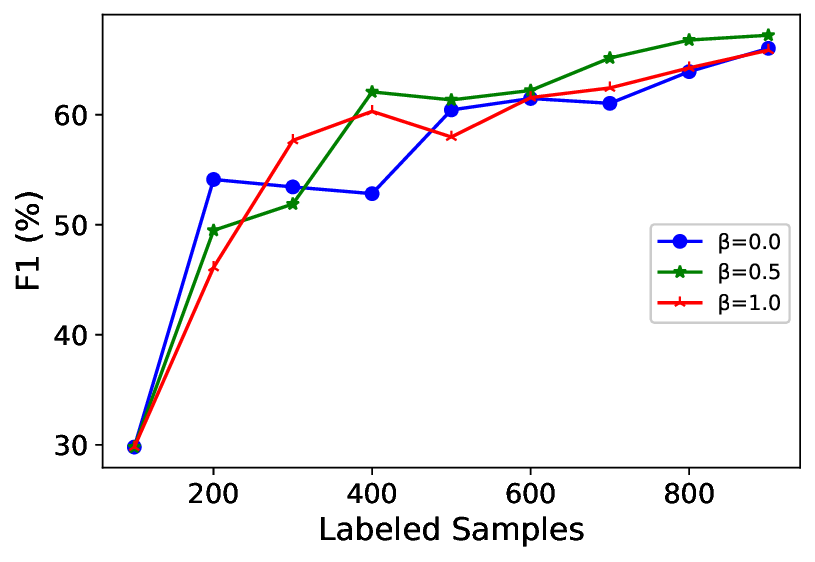}
		\caption{Amazon-Google}
		\label{fig:Amazon-Google_beta}
	\end{subfigure}
	\vspace{-2.2mm}
	\caption{F1 performance {\em vs.} cumulative number of labeled samples for different $\beta$ values. When $\beta=0$ the certainty is calculated only with respect to the spatial confidence (Eq.~\ref{eq:spatial_confidence}), and when $\beta=1$ only by the model confidence.}
	\label{fig:beta_analysis}
\end{figure}

Figure~\ref{fig:beta_analysis} displays the F1 performance as a function of the labeled set size. In both datasets the model obtained with $\beta=0.5$ performs better than the other two values when the number of labeled samples surpasses $500$. This implies that the fusion between local confidence score, assigned by the model, and spatial confidence, calculated with respect node's neighbors in the graph (Eq.~\ref{eq:spatial_confidence}) captures better model certainty.
For Walmart-Amazon, the fused version ($\beta=0.5$) reaches an F1 score of $79.76$, while for $\beta=0$ and $\beta=1$ scores are at $76.37$ and $77.59$, respectively. For Amazon-Google, with $\beta=0.5$ training ends with F1 of $67.23$, outperforming the F1 scores of $66.04$ (for $\beta=0$) and $65.87$ (for $\beta=1$). 

Both $\beta=0$ and $\beta=1$, although addressing a single aspect each, are still preferable over DAL (Figure~\ref{fig:F1vslabeled}), which is the best baseline. An interesting observation is that the models trained with $\beta=0$ and $\beta=1$ tend to fluctuate more often than the fused version, demonstrating the robustness of the fused approach. 

\subsection{Criteria Weighting Analysis}
\label{sec:alpha_analysis}
The sample selection mechanism relies, alongside correspondence, on a weighted ranking of certainty (Section~\ref{sec:Certainty}) and centrality (Section~\ref{sec:Centrality}), using the balancing factor $\alpha$ (see Eq.~\ref{eq:weighted_rank}). In addition to three different $\alpha$ values ($0.25$, $0.5$ and $0.75$), which offer different combinations of both criteria, we evaluated the performance of two sub-versions of the battleship approach, namely \emph{Battleship (cen)} and \emph{Battleship (unc)}, with $\alpha=0.0$ and $\alpha=1.0$, respectively (see Section~\ref{sec:baseline_methods}). For all experiments, and following Section~\ref{sec:beta_analysis}, we fixed $\beta$ at $0.5$.

\begin{table}[htpb]
	\centering
	\scalebox{.87}{\begin{tabular}{|l|c|c|c|c|c|}
			\hline
			\textbf{Dataset} & \textbf{$\alpha=0.0$} & \textbf{$\alpha=0.25$}   & \textbf{$\alpha=0.5$}  & \textbf{$\alpha=0.75$} & \textbf{$\alpha=1.0$}  \\\hline
			\multirow{1}{*}{\begin{tabular}[c]{@{}c@{}}Walmart-Amazon\end{tabular}} 
			& 77.71  & 78.04 & \textbf{79.76} & 76.14 & 76.13 \\\hline
			\multirow{1}{*}{\begin{tabular}[c]{@{}c@{}}Amazon-Google\end{tabular}} 
			& 65.1       & 65.38 & 67.23 & \textbf{68.22} & 66.10 \\\hline
			\multirow{1}{*}{\begin{tabular}[c]{@{}c@{}}WDC Cameras\end{tabular}} 
			& 83.85   &\textbf{86.53} & 84.97 & 82.79 & 82.22  \\\hline
			\multirow{1}{*}{\begin{tabular}[c]{@{}c@{}}WDC Shoes\end{tabular}} 
			& 66.08   & 68.48 & 72.98 & \textbf{73.24} & 71.65  \\\hline
			\multirow{1}{*}{\begin{tabular}[c]{@{}c@{}}ABT-Buy\end{tabular}} 
			& 83.21    & 86.07 & 84.31 & \textbf{87.59} & 81.52  \\\hline
			\multirow{1}{*}{\begin{tabular}[c]{@{}c@{}}DBLP-Scholar\end{tabular}} 
			& 93.95   & 94.47 & \textbf{96.03} & 93.75 & 93.81  \\\hline
	\end{tabular}}
	\caption{Analysis with various $\alpha$ values.}
	\label{tab:alpha_analysis_table}
\end{table}

Table~\ref{tab:alpha_analysis_table} displays the F1 average scores (after the final iteration) for the different datasets using the various $\alpha$ values.
For all datasets, the model has the best performance for $\alpha \in \left\lbrace 0.25, 0.5, 0.75\right\rbrace $, \emph{i.e.}, taking into account both criteria. This implies that both local centrality and model's confidence are important factors in identifying informative samples that can complement the model training.
It is noteworthy that in most cases, the results obtained with $\alpha \in \left\lbrace 0, 1\right\rbrace $  outperforms DAL, the most competitive active learning baseline. This suggests that the correspondence criterion, handled by the graph structure and budget distribution, also plays a key role in the sample selection mechanism, as discussed in Section~\ref{sec:correspondence_effect}.

\begin{figure}[b]
	\vspace{2mm}
	\centering
	\begin{subfigure}[b]{0.72\columnwidth}
		\centering
		\includegraphics[width=\columnwidth]{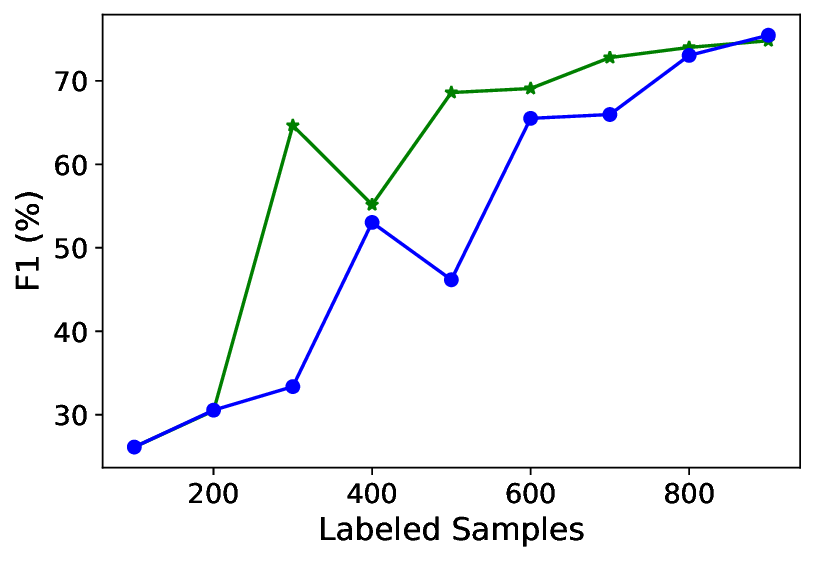}
		\caption{Walmart-Amazon}
		\label{fig:Walmart-Amazon_beta_fixed}
	\end{subfigure}
	\hfill
	\begin{subfigure}[b]{0.72\columnwidth}
		\centering
		\includegraphics[width=\columnwidth]{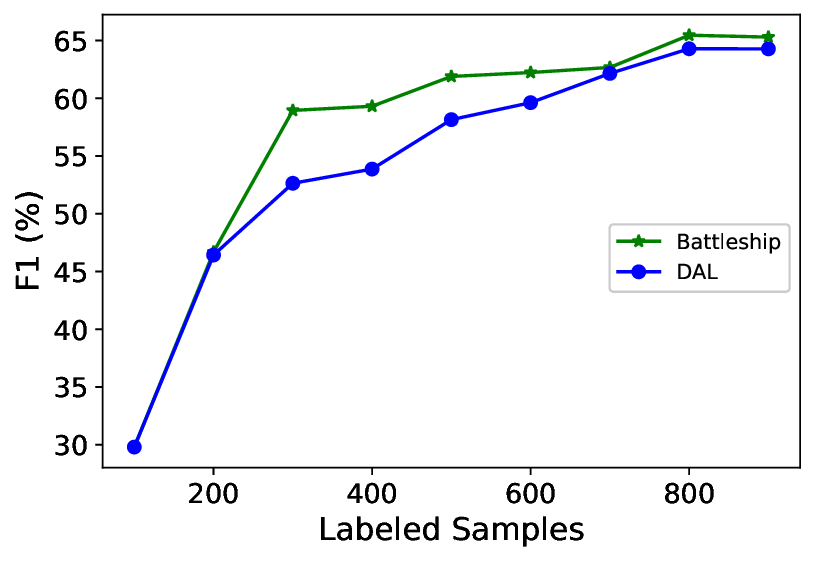}
		\caption{Amazon-Google}
		\label{fig:Amazon-Google_beta_fixed}
	\end{subfigure}
	\vspace{-2mm}
	\caption{F1 performance {\em vs.} cumulative number of labeled samples for $\alpha=1, \beta=1$.}
	\label{fig:beta_analysis_fixed}
\end{figure}

\subsection{The Correspondence Effect}
\label{sec:correspondence_effect}
To quantify the contribution of the correspondence criterion, although not directly affected by Eq.~\ref{eq:uncertainty_score}, we fixed the values $\alpha=1$ and $\beta=1$. In this case, the certainty score is determined only by the confidence assigned by the model, and the final ranking is performed only according to the certainty criterion (Eq.~\ref{eq:weighted_rank}). These conditions dictate the same selection mechanism as the one proposed by DAL, with the exception that in our approach samples are confined to their connected component (Section~\ref{sec:graphs}). In other words, with these parameters the selection of DAL is used over connected component.

\update{Figure~\ref{fig:beta_analysis_fixed} provides evidence of the effectiveness of vector-space partitioning and budget distribution policy. In the case of the Amazon-Google dataset, the battleship approach consistently outperforms DAL throughout the active learning process. For the Walmart-Amazon dataset, the battleship approach achieves an F1 value of $74.81$ after 8 active learning iterations, slightly lower than DAL's ($75.47$). However, the battleship approach demonstrates notably superior performance in terms of AUC scores (485.20 vs 418.46). This is attributed to the fact that the battleship approach consistently leads over DAL in previous iterations, with DAL only surpassing in the final iteration.}

\begin{figure}[t]
	\centering
	\begin{subfigure}[b]{0.72\columnwidth}
		\centering
		\includegraphics[width=\columnwidth]{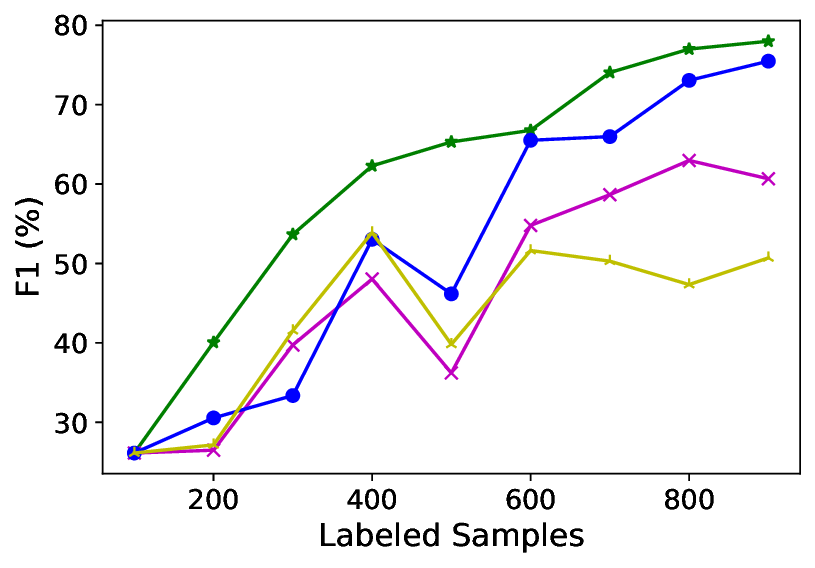}
		\caption{Walmart-Amazon}
		\label{fig:Walmart-Amazon_WS_VS_NO_WS}
	\end{subfigure}
	\hfill
	\begin{subfigure}[b]{0.72\columnwidth}
		\centering
		\includegraphics[width=\columnwidth]{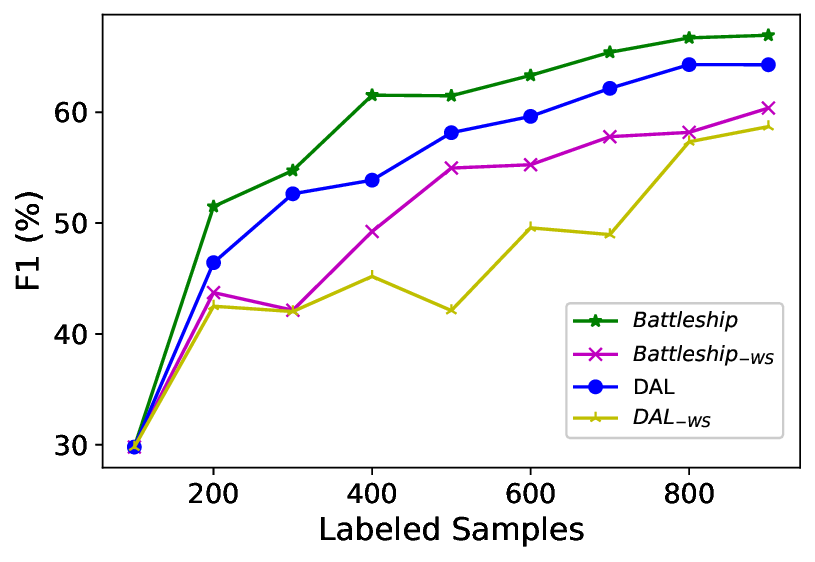}
		\caption{Amazon-Google}
		\label{fig:Amazon-Google_WS_VS_NO_WS}
	\end{subfigure}
	\vspace{-2mm}
	\caption{F1 scores with and without weak supervision.}
	\label{fig:WS_VS_NO_WS}
\end{figure}

\subsection{Weak Supervision Impact}
\label{sec:ws_analysis}
\update{We conducted a comprehensive examination to investigate the contribution of the weak supervision component (Section~\ref{sec:weak_supervision}) on our proposed battleship approach. Specifically, we assess the effectiveness of incorporating weak supervision within both the battleship approach and DAL. Furthermore, we compare our weak supervision approach, which identifies the most confident pairs based on the principles outlined in Section~\ref{sec:Certainty}, with the weak supervision approach employed in the baseline method DAL.}

\subsubsection{Weak Supervision Impact}
\label{sec:ws_and_without_ws}
\update{Figure~\ref{fig:WS_VS_NO_WS} shows the performance of the battleship approach (average scores computed over $\left\lbrace 0.25, 0.5, 0.75\right\rbrace $, as reported in Section~\ref{sec:prog_f1}) and DAL with and without weak supervision ($Battleship$/$DAL$ and $Battleship_{-WS}$/$DAL_{-WS}$, respectively).} 

\update{For the Walmart-Amazon dataset, incorporating weak supervision into the battleship approach ($Battleship_{ws}$) yields progressively higher scores as the number of labeled samples increases, peaking at 77.98. Conversely, the battleship approach without weak supervision ($Battleship_{-WS}$) demonstrates variable performance, with a maximum score of 60.66. Likewise, in the case of DAL, removing weak supervision ($DAL_{-WS}$) results in decreased scores, reaching a maximum of 50.7.
A similar pattern is observed in the Amazon-Google dataset, where the inclusion of weak supervision leads to consistent improvement within the battleship approach, achieving a maximum score of 66.94. In contrast, $Battleship_{-WS}$ exhibits more fluctuation, with a maximum score of 60.37. Similarly, DAL demonstrates a comparable trend, with weak supervision contributing to steady performance enhancement (64.28), while its absence leads to more variable results (58.7).}

\update{These results imply that 
	weak supervision is effective in enhancing the performance of both the battleship approach and DAL, leading to more stable results over learning course.}

\begin{figure}[htpb]
	\centering
	\begin{subfigure}[b]{0.72\columnwidth}
		\centering
		\includegraphics[width=\columnwidth]{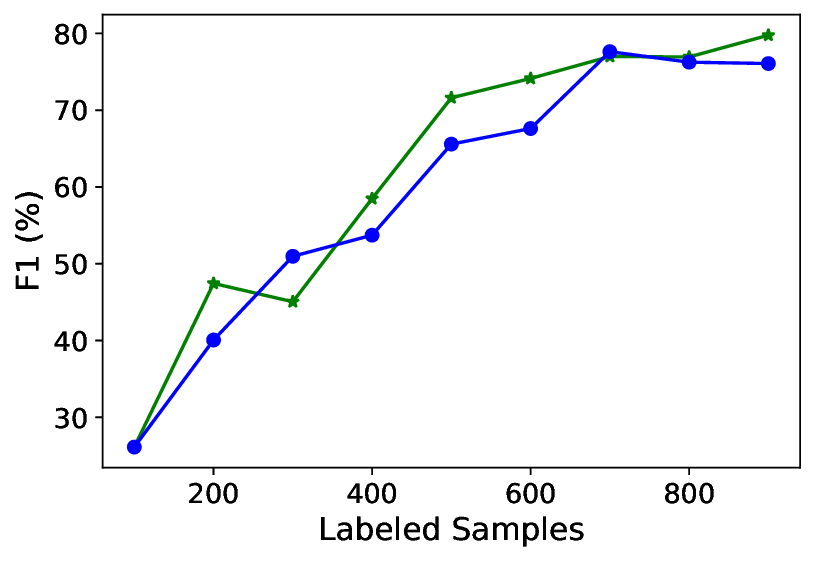}
		\caption{Walmart-Amazon}
		\label{fig:Walmart-Amazon_WS_b_WS_K}
	\end{subfigure}
	\hfill
	\begin{subfigure}[b]{0.72\columnwidth}
		\centering
		\includegraphics[width=\columnwidth]{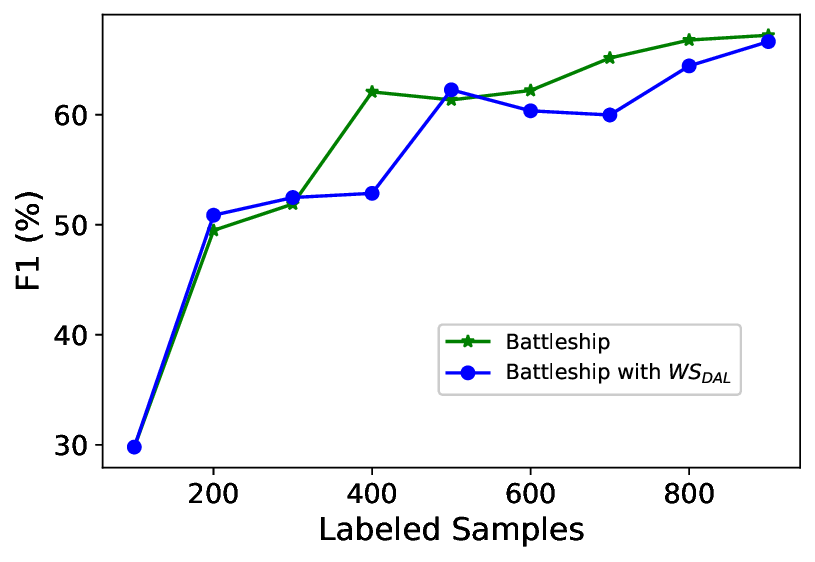}
		\caption{Amazon-Google}
		\label{fig:Amazon-Google_WS_b_WS_K}
	\end{subfigure}
	\vspace{-1.3mm}
	\caption{F1 performance of the battleship approach and the battleship approach with DAL's weak supervision.}
	\label{fig:WS_b_VS_WS_k}
\end{figure}

\subsubsection{Weak Supervision Method Comparison}
\label{sec:WS_b_VS_WS_k}
\update{As highlighted in Section~\ref{sec:weak_supervision}, DAL utilizes a weak supervision method that focuses on identifying pairs where the model exhibits high confidence, as indicated by minimizing Eq.~\ref{eq:conditional_entropy}. In our battleship approach we integrate between the model's standard conditional entropy and spatial entropy using Eq.~\ref{eq:uncertainty_score}.
Figure~\ref{fig:WS_b_VS_WS_k} presents the performance of the battleship approach (with $\alpha=0.5, \beta=0.5$) over two cases, differing only in their weak supervision method, such that $battleship$ uses Eq.~\ref{eq:uncertainty_score}, while $battleship$ with $WS_{DAL}$ employs Eq.~\ref{eq:conditional_entropy}.}

\update{For both datasets the selection mechanism of the active learning process is identical, yet, it can be seen that when using the weak supervision method of the battleship approach the obtained results are slightly better than when relying only the model's confidence. For Amazon-Google, the battleship approaches reaches an AUC score of 467.49, outperforming $battleship$ with $WS_{DAL}$ (451.49). Similarly, for Walmart-Amazon the $battleship$ approaches also beats its baseline (503.58 and 482.92, respectively).
Based on the results, it can be observed that both approaches demonstrate competitive performance. However, the subtle variations in scores emphasize the advantage of the battleship approach's weak supervision method.}

\subsection{Analysis Conclusion}
\label{sec:ablation_conclusion}
The parameters $\alpha$ and $\beta$ are central components of our approach. Anyhow, there is no single setting that is consistently better. A noteworthy observation from our experiments is that incorporating a combination of components (centrality vs. uncertainty and local vs. spatial certainty) yields better performance compared to disregarding a component entirely (\emph{i.e.}, $\alpha \in \left\lbrace 0,1 \right\rbrace $, $\beta \in \left\lbrace 0,1 \right\rbrace $). This highlights the challenge of determining optimal values for both alpha and beta in advance, as their interplay and combined effects are crucial for achieving favorable results. Further investigation into the intricate relationship between alpha and beta values may provide valuable insights for future research in this area.



%% file: related.tex
\section{Related Work}
\label{sec:related}
Active learning refers to an iterative process of selecting data samples to be labeled, which are then used to train a model. The underlying hypothesis is that with a careful selection of informative samples, a classifier can perform as well or even better than when being trained on a fully labeled dataset~\cite{settles2009active}.
Broad range of active learning approaches to the entity matching task have been proposed over the last two decades~\cite{tejada2001learning, sarawagi2002interactive, qian2017active}. By and large, they can be divided into three main groups, namely rule extraction, traditional machine learning, and deep learning.

Qian {\em et al.}~\cite{qian2017active} presented a large-scale system of rule learning to improve recall. They focus on identifying atomic matching predicate combinations (operations such as equality and similarity over attributes) that are relevant to identifying matched pairs. Isele and Bizer~\cite{isele2013active} offer an interactive rule generation method, with the aim of minimizing user involvement in the process.
Meduri {\em et al.}~\cite{meduri2020comprehensive} surveys different sample selection approaches, examining combinations of different approaches alongside combinations of classifiers~\cite{tejada2001learning, sarawagi2002interactive}. A large share of this survey~\cite{meduri2020comprehensive} concentrates on the concept of Query-by-committee (QBC)~\cite{freund1997selective, seung1992query} as a policy for sample selection during the active learning process. Typically, 
QBC finds uncertain samples (which are believed to be more informative to the model) by training multiple versions of a classifier and measuring uncertainty as their level of disagreement. For example, Mozafari \emph{et al.}~\cite{mozafari2014scaling} define the variance of the committee for the matching task as $X\left( u\right) \left( 1 -  X\left( u\right) \right)$ where $X\left( u\right)$ is the fraction of classifiers predicted that a given pair is a match.

Deep learning has become the dominant approach to entity matching tasks. Ebraheem {\em et al.}~\cite{joty2018distributed} were the firsts to apply neural networks to entity matching, followed by Mudgal {\em et al.}~\cite{mudgal2018deep}, which introduced a design space for the use of deep learning to this task. Recently, several work used pre-trained language models to tackle the entity matching problem~\cite{li2020deep, brunner2020entity, li2021deep, peeters2021dual}. In our work, we use DITTO~\cite{li2020deep} as a major component of the battleship, training it as a matcher after each active learning iteration, and using it to produce \tup pair representations.

Several recent works~\cite{kasai2019low, jain2021deep, thirumuruganathan2018reuse, bogatu2021cost} use deep learning to tackle the entity matching problem under active learning settings. Kasai {\em et al.}~\cite{kasai2019low} trains a deep learning-based transferable model, aimed at domain prediction task. This model is integrated with active learning process, looking for informative samples to enrich the transferable model in following iterations. 
Thirumuruganathan {\em et al.}~\cite{thirumuruganathan2018reuse} treat vector-space representations of \tups as features and apply traditional machine learning algorithms to adapt a classifier from one domain to another. Bogatu {\em et al.}~\cite{bogatu2021cost} use variational auto-encoders for generating entity representations, then utilize them to create a transferable model. We do not use any designated architecture for domain adaptation, assuming that labeled data from a source domain is not available.

Jain {\em et al.}~\cite{jain2021deep} use transformer pre-trained language model as a classifier, training conjointly blocker and matcher. For blocking, they obtain a vector-space representation to \tups, and run a similarity search to find potential matches, while for matching they use DITTO~\cite{li2020deep}. 
Unlike these aforementioned works, we utilize \tup pair representations (instead of single \tup representation) as part of the sample selection, instead of using it only for prediction~\cite{li2020deep}. In addition, we expand the notion of uncertainty, which serves as a sample selection criterion~\cite{kasai2019low, jain2021deep}, allowing spatial considerations to be taken into account. \update{Beyond the scope of the entity matching task, Mahdavi et al.~\cite{mahdavi2019raha} has also utilized sample representations for sample diversity, addressing the error detection task. They based their selection on a feature vector derived from multiple error detection strategies, then employed clustering and label propagation to select representative data samples.}

Another approach for dealing with the lack of labeled data was suggested by Wu {\em et al.}~\cite{wu2020zeroer}. Their solution, termed ZeroER, does not rely on labeled data at all, but on the assumption that feature vectors of matching pairs (built upon variety of similarity measures) are distributed in a different way than those of non-matching pairs. In our work we use an increasing training set to obtain pair representations (as an alternative to model-agnostic feature vector) and use them to find the most informative samples to label in the following iteration.

%% file: conclusions.tex
\vspace{-1.5mm}
\section{Conclusions}
\label{sec:conclusions}
In this work we introduce the battleship approach, a novel active learning method for solving the entity matching problem.
The approach uses \tup pair representations, utilizing spatial (vector-space) considerations to spot informative data samples that are labeling-worthy. We tailor sample selection to the characteristics of the entity matching problem, establishing the intuition and motivation for the various design choices made in our algorithm. A thorough empirical evaluation shows that the battleship approach is an effective solution to entity matching under low-resource conditions, in some cases even more than a model that was trained against data without labeling budget limitation. In future work we aspire to expand the battleship approach beyond the 
entity matching, {\em e.g.}, to Natural Language Processing tasks, as we believe that some of the main principles that guided us in this work can be generalized and applied to a broader range of 
challenges.

%% file: acknowledgments.tex
\begin{acks}
This work was supported in part by the National Science Foundation (NSF) under award numbers IIS-1956096. We also acknowledge the support of the Benjamin and Florence Free Chair.
\end{acks}